\documentstyle[preprint,aps,eqsecnum]{revtex} 
\begin{document}



\def\YM/{Yang\discretionary{-}{}{-}Mills}
\def\FT/{Freedman\discretionary{-}{}{-}Townsend}
\def\CS/{Chern\discretionary{-}{}{-}Simons}
\def\YMH/{Yang\discretionary{-}{}{-}Mills\discretionary{-}{}{-}Higgs}
\def\EL/{Euler\discretionary{-}{}{-}Lagrange}

\hyphenation{
all
along
anti
ap-pen-dix 
co-tan-gent
equa-tion
equa-tion-s
equiv-a-lent
evo-lu-tion
fields
form
iden-ti-ty
iden-ti-ties
im-por-tant
its
La-grang-ian
La-grang-ian-s
next
nev-er
prod-uct
real
sca-lar
sym-me-try
sym-me-tries
tak-en
tan-gent
term
two
use
use-s
vari-a-tion
vari-a-tion-s
}


\def\Rnum{{\bf R}}
\def\Cnum{{\bf C}}

\def\eqref#1{(\ref{#1})}
\def\eqrefs#1#2{(\ref{#1}) and~(\ref{#2})}
\def\eqsref#1#2{(\ref{#1}) to~(\ref{#2})}

\def\Eqref#1{Eq.~(\ref{#1})}
\def\Eqrefs#1#2{Eqs.~(\ref{#1}) and~(\ref{#2})}
\def\Eqsref#1#2{Eqs.~(\ref{#1}) to~(\ref{#2})}

\def\secref#1{Sec.~\ref{#1}}
\def\secrefs#1#2{Secs.~\ref{#1} and~\ref{#2}}

\def\appref#1{App.~\ref{#1}}

\def\Ref#1{Ref.~\cite{#1}}

\def\Cite#1{${\mathstrut}^{\cite{#1}}$}

\def\tableref#1{Table~\ref{#1}}

\def\figref#1{Fig.~\ref{#1}}

\hyphenation{Eq Eqs Sec App Ref Fig}

\def\EQ{\begin{equation}}
\def\EQs{\begin{eqnarray}}
\def\endEQ{\end{equation}}
\def\endEQs{\end{eqnarray}}

\def\proclaim#1{\medbreak
\noindent{\it {#1}}}
\def\Proclaim#1#2{\medbreak
\noindent{\bf {#1}}{\it {#2}}\par\medbreak}


\def\newline{\hfil\break}
\def\fewquad{\qquad\qquad}
\def\severalquad{\qquad\fewquad}
\def\manyquad{\qquad\severalquad}
\def\manymanyquad{\manyquad\manyquad}

\def\mstrut{\mathstrut}
\def\hp#1{\hphantom{#1}}

\def\ontop#1#2{
\setbox2=\hbox{{$#2$}} \setbox1=\hbox{{$\scriptscriptstyle #1$}} 
\dimen1=0.5\wd2 \advance\dimen1 by 0.5\wd1 \dimen2=1.4\ht2
\ifdim\wd1>\wd2 \raise\dimen2\box1 \kern-\dimen1 \hbox to\dimen1{\box2\hfill}
\else \box2\kern-\dimen1 \raise\dimen2 \hbox to\dimen1{\box1\hfill} \fi }

\def\mixedindices#1#2{{\mstrut}^{\mstrut #1}_{\mstrut #2}}
\def\downindex#1{{\mstrut}^{\mstrut}_{\mstrut #1}}
\def\upindex#1{{\mstrut}_{\mstrut}^{\mstrut #1}}
\def\downupindices#1#2{{\mstrut}_{\mstrut #1}^{\hp{#1}\mstrut #2}}
\def\updownindices#1#2{{\mstrut}^{\mstrut #1}_{\hp{#1}\mstrut #2}}

\def\tensor#1#2#3{{#1}\mixedindices{#2}{#3}}
\def\covector#1#2{{#1}\downindex{#2}}
\def\vector#1#2{{#1}\upindex{#2}}

\def\id#1#2{\delta\downupindices{#1}{#2}}
\def\cross#1#2{\epsilon\,\downupindices{#1}{#2}}
\def\vol#1{\epsilon\,\downindex{#1}}
\def\invvol#1{\epsilon\,\upindex{#1}}
\def\x#1{x\upindex{#1}}

\def\metric#1{g\downindex{#1}}
\def\invmetric#1{g\upindex{#1}}
\def\flatmetric#1{\eta\downindex{#1}}
\def\invflatmetric#1{\eta\upindex{#1}}

\def\matr#1{\left( \matrix{#1} \right)}

\def\A#1#2{\tensor{A}{#1}{#2}}
\def\F#1#2{\tensor{F}{#1}{#2}}
\def\stF#1#2{\tensor{\tilde F}{#1}{#2}}
\def\B#1#2{\tensor{B}{#1}{#2}}
\def\H#1#2{\tensor{H}{#1}{#2}}
\def\stH#1#2{\tensor{\tilde H}{#1}{#2}}

\def\sX#1{\vector{\xi}{#1}}
\def\vX#1#2{\tensor{\chi}{#1}{#2}}
\def\sXvar{\delta_\xi}
\def\vXvar{\delta_\chi}
\def\sXsub#1#2{\vector{\xi_{#1}}{#2}}
\def\vXsub#1#2#3{\tensor{\chi_{#1}}{#2}{#3}}
\def\sXvarsub#1{\delta_{\xi_{#1}}}
\def\vXvarsub#1{\delta_{\chi_{#1}}}

\def\EA#1#2{\tensor{E_{\rm A}}{#1}{#2}}
\def\EB#1#2{\tensor{E_{\rm B}}{#1}{#2}}
\def\coEA#1#2{\tensor{E^{\rm A}}{#2}{#1}}
\def\coEB#1#2{\tensor{E^{\rm B}}{#2}{#1}}

\def\striv{\Theta_{\sX{}}}
\def\vtriv{\Theta_{\vX{}{}}}

\def\Y#1#2{Y\mixedindices{#1}{#2}}
\def\invY#1#2{Y\inv\mixedindices{#1}{#2}}

\def\K#1#2{\tensor{K}{#1}{#2}}
\def\P#1#2{\tensor{P}{#1}{#2}}
\def\Q#1#2{\tensor{Q}{#1}{#2}}
\def\stK#1#2{\tensor{\tilde K}{#1}{#2}}
\def\stP#1#2{\tensor{\tilde P}{#1}{#2}}
\def\stQ#1#2{\tensor{\tilde Q}{#1}{#2}}

\def\charge#1#2{\vector{Q_{\rm #1}}{#2}}
\def\J#1#2#3{\tensor{J_{\rm #1}}{#2}{#3}}

\def\T#1{T\downindex{#1}}

\def\kv#1{\vector{\zeta}{#1}}
\def\t#1#2{\tensor{t}{#1}{#2}}

\def\parity{{\cal P}}

\def\der#1{\partial\downindex{#1}}
\def\coder#1{\partial\upindex{#1}}

\def\D#1{D\downindex{#1}}
\def\coD#1{D\upindex{#1}}

\def\Parder#1#2{
\mathchoice{\partial{#1} \over\partial{#2}}{\partial{#1}/\partial{#2}}{}{} }
\def\parder#1{\partial/\partial{#1}}

\def\DA#1{D_{#1}}
\def\DB#1{D'_{#1}}

\def\Lie#1{{\cal L}{}_{#1}}

\def\Ader#1{\der{\A{#1}{}}}
\def\Bder#1{\der{\B{#1}{}}}
\def\dAder#1{\der{d\A{#1}{}}}
\def\dBder#1{\der{d\B{#1}{}}}
\def\kthdAder#1{\der{(d*)^k d\A{#1}{}}}
\def\kthdBder#1{\der{(d*)^k d\B{#1}{}}}
\def\vec#1#2{f\mixedindices{#1}{\rm #2}}
\def\covec#1#2{f\mixedindices{\rm #2}{#1}}

\def\Acoder#1{\partial\mixedindices{\flat}{\A{#1}{}}}
\def\Bcoder#1{\partial\mixedindices{\flat}{\B{#1}{}}}
\def\kthdAcoder#1{\partial\mixedindices{\flat}{(d*)^k d\A{#1}{}}}
\def\kthdBcoder#1{\partial\mixedindices{\flat}{(d*)^k d\B{#1}{}}}

\def\ELop#1{{\rm E}\mixedindices{\flat}{#1}}
\def\coELop#1{{\rm E}\mixedindices{\sharp}{#1}}

\def\ymTD#1{D_{\rm #1}^{\rm T}}
\def\ymD#1{D_{\rm #1}}
\def\curvA#1{R_{#1}}
\def\curvB#1{R'_{#1}}
\def\conxA#1{\Gamma_{#1}}
\def\conxB#1{\Gamma'_{#1}}

\def\ymA#1{{\rm A_{#1}}}
\def\ymF#1{{\rm F_{#1}}}
\def\ftB#1{{\rm B_{#1}}}
\def\ftK#1{{\rm K_{#1}}}
\def\ftY#1{{\rm Y_{#1}}}
\def\ftYinv#1{{\rm Y}_{\rm #1}^{-1}}

\def\e#1{e\downindex{#1}}
\def\idmap{{\rm id}}

\def\ymLB#1#2{\varepsilon\updownindices{#1}{#2}}
\def\ftLB#1#2{c\updownindices{#1}{#2}}
\def\ymTLB#1#2{\varepsilon\downupindices{#1}{#2}}
\def\ftTLB#1#2{c\downupindices{#1}{#2}}
\def\ymCK#1#2{k\downupindices{#1}{#2}}
\def\auxLB#1#2{f\downupindices{#1}{#2}}
\def\auxTLB#1#2{f\updownindices{#1}{#2}}

\def\vsmap#1#2{h\downupindices{#1}{#2}}

\def\u#1#2{\tensor{u}{#1}{#2}}
\def\v#1#2{\tensor{v}{#1}{#2}}
\def\w#1#2{\tensor{w}{#1}{#2}}
\def\uB#1#2{\tensor{u'}{#1}{#2}}
\def\vB#1#2{\tensor{v'}{#1}{#2}}
\def\wB#1#2{\tensor{w'}{#1}{#2}}

\def\SU#1{SU(#1)}
\def\SO#1{SO(#1)}
\def\U#1{U(#1)}
\def\G{{\cal G}}

\def\vsA{{\cal A}}
\def\vsB{{\cal A'}}

\def\adA{ad_\vsA}
\def\adB{ad_\vsB}
\def\auxAh{\adA\circ\BintoA}
\def\auxhA{ad_{\BintoA,\vsA}}
\def\adTA{ad_\vsA^{\rm T}}
\def\adTB{ad_\vsB^{\rm T}}
\def\adcoTA{ad_\vsA^*}
\def\adcoTB{ad_\vsB^*}

\def\auxTAh{\adTA\circ\BintoA}
\def\auxThA{ad_{\BintoA,\vsA}^{\rm T}}
\def\auxcoThA{ad_{\BintoA,\vsA}^*}
\def\auxBh{\adB\circ\TBintoA}
\def\auxhB{ad_{\BintoA,\vsB}}
\def\auxTBh{\adB^{\rm T}\circ\TBintoA}
\def\auxTBhinv{\adB^{\rm T}\circ\BintoA\inv}
\def\auxThB{ad_{\BintoA,\vsB}^{\rm T}}
\def\auxcoThB{ad_{\BintoA,\vsB}^*}

\def\bilin{f}
\def\BintoA{h}
\def\TBintoA{h^{\rm T}}
\def\Tbilin{f^{\rm T}}

\def\m#1#2{m\downupindices{#1}{#2}}
\def\Tm#1#2{m\downupindices{#1}{#2}}
\def\mA{m_\vsA}
\def\mB{m_\vsB}
\def\minvAB{m_\vsB^{-1}}

\def\adAzero{ad_{\vsA_0}}
\def\adAm{ad_{\vsA_{\rm m}}}

\def\projA{{\cal P}_0}
\def\mprojA{{\cal P}_{\rm m}}
\def\projB{{\cal P}'_0}
\def\mprojB{{\cal P}'_{\rm m}}

\def\reprA{\rho}
\def\reprB{\rho'}
\def\reprAm{\rho_{\rm m}}
\def\reprBm{\rho'_{\rm m}}
\def\treprAm{{\tilde \rho}_{\rm m}}

\def\semi{ \setbox1=\hbox{{$\times$}} \setbox2=\hbox{{$\scriptstyle |$}}
\box1\kern -1.3\wd2 \raise.2\ht2\box2 }

\def\intprod{ {\scriptstyle\rfloor} }

\def\c{\case{1}{\sqrt 2} \frac{\kappa}{m}}

\def\fieldsp{{\cal S}}
\def\jetsp{J^{(\infty)}(\fieldsp)}


\def\linF#1{\vector{F}{#1}}
\def\linH#1{\vector{H}{#1}}
\def\linstF#1{\vector{\tilde F}{#1}}
\def\linstH#1{\vector{\tilde H}{#1}}

\def\nthEA#1#2{\vector{\ontop{(#1)}{E_{\rm A}}}{#2}}
\def\nthEB#1#2{\vector{\ontop{(#1)}{E_{\rm B}}}{#2}}
\def\nthcoEA#1#2{\covector{\ontop{(#1)}{E^{\rm A}}}{#2}}
\def\nthcoEB#1#2{\covector{\ontop{(#1)}{E^{\rm B}}}{#2}}
\def\nthL#1{\ontop{(#1)}{L}}

\def\nthsXvar#1{\ontop{(#1)}{\delta_\xi}}
\def\nthvXvar#1{\ontop{(#1)}{\delta_\chi}}
\def\nthcommXvar#1#2#3{\ontop{(#1)}{[\delta_{#2},\delta_{#3}]}}
\def\nthsXvarsub#1#2{\ontop{(#1)}{\delta_{\xi_{#2}}}}
\def\nthvXvarsub#1#2{\ontop{(#1)}{\delta_{\chi_{#2}}}}

\def\nthstriv#1{\ontop{(#1)}{\Theta_\xi}}
\def\nthvtriv#1{\ontop{(#1)}{\Theta_\chi}}
\def\nthsX#1#2{\ontop{(#1)}{\sX{#2}}}
\def\nthvX#1#2#3{\ontop{(#1)}{\vX{#2}{#3}}}
\def\nthsXsub#1#2#3{\ontop{(#1)}{\xi}{}^{#2}_{#3}}
\def\nthvXsub#1#2#3{\ontop{(#1)}{\mstrut\chi}{}^{#2}_{#3}}

\def\nthstriv#1{\ontop{(#1)}{\Theta_{\sX{}}}}
\def\nthvtriv#1{\ontop{(#1)}{\Theta_{\vX{}{}}}}

\def\atens#1#2{a\upindex{#1}\downindex{#2}}
\def\btens#1#2{b\upindex{#1}\downindex{#2}}
\def\ctens#1#2{c\upindex{#1}\downindex{#2}}
\def\tctens#1#2{{\tilde c}\upindex{#1}\downindex{#2}}
\def\dtens#1#2{d\upindex{#1}\downindex{#2}}
\def\tdtens#1#2{{\tilde d}\upindex{#1}\downindex{#2}}
\def\etens#1#2{e\upindex{#1}\downindex{#2}}
\def\tetens#1#2{{\tilde e}\upindex{#1}\downindex{#2}}
\def\ftens#1#2{f\upindex{#1}\downindex{#2}}
\def\gtens#1#2{g\upindex{#1}\downindex{#2}}
\def\tgtens#1#2{{\tilde g}\upindex{#1}\downindex{#2}}
\def\htens#1#2{h\upindex{#1}\downindex{#2}}
\def\thtens#1#2{{\tilde h}\upindex{#1}\downindex{#2}}
\def\itens#1#2{i\upindex{#1}\downindex{#2}}
\def\titens#1#2{{\tilde i}\upindex{#1}\downindex{#2}}
\def\jtens#1#2{j\upindex{#1}\downindex{#2}}
\def\tjtens#1#2{{\tilde j}\upindex{#1}\downindex{#2}}
\def\ktens#1#2{k\upindex{#1}\downindex{#2}}
\def\tktens#1#2{{\tilde k}\upindex{#1}\downindex{#2}}
\def\ltens#1#2{l\upindex{#1}\downindex{#2}}
\def\tltens#1#2{{\tilde l}\upindex{#1}\downindex{#2}}
\def\mtens#1#2{m\upindex{#1}\downindex{#2}}
\def\tmtens#1#2{{\tilde m}\upindex{#1}\downindex{#2}}

\def\Tatens#1#2{a\downindex{#1}\upindex{#2}}
\def\Tktens#1#2{k\downindex{#1}\upindex{#2}}
\def\Tjtens#1#2#3{j\downindex{#1}\upindex{#2}\downindex{#3}}
\def\Tbtens#1#2#3{b\downindex{#1}\upindex{#2}\downindex{#3}}
\def\Tetens#1#2{e\downindex{#1}\upindex{#2}}

\def\frac#1#2{{\textstyle {#1 \over #2}}}

\def\Rnum{\bold R}
\def\Cnum{\bold C}

\def\inv{{}^{-1}}
\def\ad{{}^{\rm T}}

\def\const{{\rm const.}}

\def\ie/{i.e.}
\def\eg/{e.g.}


\title{ Gauge theories of Yang-Mills vector fields coupled to 
antisymmetric tensor fields }

\author{Stephen C. Anco\cite{email}}
\address{
Department of Mathematics\\ 
Brock University
St Catharines, ON Canada L2S 3A1\\
}

\maketitle

\begin{abstract}
A nonabelian class of massless/massive nonlinear gauge theories of 
\YM/ vector potentials coupled to \FT/ antisymmetric tensor potentials
is constructed in four spacetime dimensions. 
These theories involve an extended \FT/ type coupling 
between the vector and tensor fields, 
and a \CS/ mass term with the addition of a Higgs type coupling
of the tensor fields to the vector fields in the massive case. 
Geometrical, field theoretic, and algebraic aspects of 
the theories are discussed in detail. 
In particular, 
the geometrical structure mixes and unifies features of 
\YM/ theory and \FT/ theory formulated 
in terms of Lie algebra valued curvatures and connections associated to
the fields and nonlinear field strengths. 
The theories arise from a general determination of
all possible geometrical nonlinear deformations of 
linear abelian gauge theory for 1-form fields and 2-form fields
with an abelian \CS/ mass term in four dimensions. 
For this type of deformation
(with typical assumptions on the allowed form considered for
terms in the gauge symmetries and field equations),
an explicit classification of deformation terms at first-order is obtained,
and uniqueness of deformation terms at all higher-orders is proven. 
This leads to a uniqueness result for the nonabelian class of theories
constructed here. 
\end{abstract}

\narrowtext
\tightenlines
\newpage

\section{ Introduction }
\label{intro}

Gauge field theories continue to be fundamental in the study of 
many areas of mathematical physics,
ranging from elementary particle interactions, 
and completely integrable nonlinear differential equations,
to topology of three and four dimensional manifolds. 
Consequently, an effort to generalize the important types of 
gauge field theories is of natural interest. 
In recent work \cite{JMPpaper1,JMPpaper2}, 
a new nonlinear gauge theory was found 
for massless vector fields in three spacetime dimensions, 
describing a novel type of generalization of nonabelian \YM/ theory. 
Its origin can be understood by considering nonlinear deformations of
the abelian linear gauge theory of 1-form potentials in $d$ dimensions
\cite{AMSpaper,Henneaux1,BarnichBrandtHenneaux1}.

The deformation process considered here consists of 
adding linear and higher power terms to the abelian gauge symmetry
while also adding quadratic and higher power terms to 
the linear field equations, 
such that a gauge invariant action principle exists 
which is not equivalent to the undeformed linear theory 
under nonlinear field redefinitions. 
The property of gauge invariance is very restrictive 
and can be used to derive determining equations 
for the allowed form of the deformation terms added order by order
in powers of the fields. 

Nonabelian \YM/ theory describes one type of allowed deformation,
which works for 1-form potentials in any dimension $d>1$. 
Interestingly, in $d=3$ dimensions, 
another type of deformation is allowed \cite{JMPpaper1},
analogous to the \FT/ theory of antisymmetric tensor gauge fields 
\cite{FTth}.
The \FT/ theory was derived originally only 
for antisymmetric tensor fields in $d=4$ dimensions 
but it has a simple geometrical formulation in any dimension $d>2$
as a nonlinear gauge theory of $d-2$-form potentials,
in particular, 1-form potentials in $d=3$ dimensions. 
Moreover, 
this formulation of the theory has a further natural extension to 
a coupled tower of $p$-form potentials of all ranks $1\le p\le d-2$, 
in particular, coupled 1-form and 2-form potentials in $d=4$ dimensions
\cite{Henneaux2,Brandt1}.

The novel generalization of nonabelian \YM/ theory in \Ref{JMPpaper2}
arises by combining the \YM/ type and \FT/ type deformations of
the abelian linear 1-form potential gauge theory 
in $d=3$ dimensions. 
In the present paper, 
a similar nonlinear deformation of the abelian linear gauge theory of 
1-form and 2-form potentials 
in $d=4$ dimensions is studied, 
which has been announced in earlier work \cite{JMPpaper2,notes}.
The resulting nonlinear gauge theory 
generalizes both nonabelian \YM/ theory and \FT/ theory,
describing coupled massless vector and antisymmetric tensor fields
in four spacetime dimensions. 
As a main new result, 
an interesting extension of this theory to include a \CS/ type mass term
involving both the vector and antisymmetric tensor fields
is presented. 

Physically speaking, 
the field strengths in this nonlinear gauge theory together represent
coupled massive spin-one fields in the case with a \CS/ term,
and otherwise represent 
massless spin-one fields coupled to massless spin-zero fields
in the case with no \CS/ term. 
The construction and features of these two cases of the theory 
are given in \secref{masslessth} and \secref{massiveth}.
The theory has a very rich and interesting geometrical structure,
mixing and unifying features of \YM/ theory and \FT/ theory
in terms of curvatures and connections associated with 
the fields and field strengths, 
which is discussed in \secref{geometricalth}.
In \secref{deformations},
the theory is derived 
from an analysis of allowed nonlinear geometrical deformations of 
the abelian linear gauge theory of 
massless/massive sets of 1-form and 2-form potentials
in four dimensions, 
with the mass determined by a \CS/ type term. 
This analysis yields a novel nonlinear gauge theory 
for coupled massless and massive sets of
vector and antisymmetric tensor fields,
generalizing the two preceding cases of the new theory 
from \secrefs{masslessth}{massiveth}. 
Finally, some concluding remarks are made in \secref{conclude}.

\section{ Deformation of nonabelian Yang-Mills/Freedman-Townsend gauge theory }
\label{masslessth}

First consider, as a starting point, 
the formulation of nonabelian \YM/ theory and \FT/ theory 
as respective nonlinear gauge theories of massless 
vector and antisymmetric tensor fields
on four dimensional Minkowski spacetime. 
For simplicity,
the gauge groups will be taken to be three dimensional. 
Recall, in \YM/ theory, the Lie algebra underlying the gauge group
is required to be compact semisimple, 
which then fixes it here to be $\SU{2}$.
In \FT/ theory, however, no such condition arises on 
the underlying Lie algebra of the gauge group,
and thus here it can be any three dimensional nonabelian Lie algebra, $\G$.
From the classification of three dimensional Lie algebras,
it then follows that $\G$ 
either is semisimple and thus $\G=\SU{2}$, $\G=\SU{1,1}$,
or is solvable and thus $\G=\U{1} \semi\U{1}{}^2$ 
which is a semi-direct product of abelian Lie algebras 
$\U{1}$ and $\U{1}{}^2$.

To formulate \YM/ theory with an $\SU{2}$ gauge group,
introduce as the field variable a vector potential $\A{}{\mu}$
that takes values in the Lie algebra $\SU{2}$. 
Equivalently, with respect to a fixed $\SU{2}$ basis $\e{a}$, $a=1,2,3$,
the vector potential components 
$\A{}{\mu} =\A{a}{\mu} \e{a}$ 
can be regarded as a set of three ordinary vector fields $\A{a}{\mu}$
on Minkowski spacetime. 
Let $\ymLB{a}{bc}$ denote the $\SU{2}$ structure constants,
and let $\ymCK{ab}{}$ denote an $\SU{2}$ positive-definite invariant metric,
related to Killing metric by 
$\ymCK{ab}{} = -\ymLB{c}{ad} \ymLB{d}{bc}$,
and so $\ymLB{}{abc}= \ymLB{e}{bc} \ymCK{ae}{}$ 
is totally antisymmetric. 

The $\SU{2}$ \YM/ field strength is given by 
\EQ
\F{a}{\sigma\mu} = 
\der{[\sigma}\A{a}{\mu]} 
+ \case{1}{2} \ymLB{a}{bc} \A{b}{\sigma}\A{c}{\mu} . 
\label{ymF}
\endEQ
It is convenient in four dimensions to work with the dual field strength
\EQ
\stF{a}{\sigma\mu} =
\cross{\sigma\mu}{\tau\nu} \F{a}{\tau\nu} , 
\label{ymdualF}
\endEQ
which satisfies the Bianchi identity
\EQ
\coD{\sigma} \stF{a}{\sigma\mu} =0
\label{YMFid}
\endEQ
where 
\EQ\label{YMD}
\D{\sigma} =\der{\sigma} +\ymLB{a}{bc} \A{b}{\sigma}
\endEQ
is the $\SU{2}$ covariant derivative operator. 
The \YM/ Lagrangian is given by 
\EQ\label{ymL}
L_{\rm YM} = \case{1}{2} \ymCK{ab}{} \stF{a}{\sigma\mu} \stF{b}{\tau\nu} 
\invflatmetric{\sigma\tau} \invflatmetric{\mu\nu}
\endEQ
yielding the $\SU{2}$ \YM/ field equation
\EQ
\EA{a}{\tau} = 
\cross{\tau}{\sigma\nu\mu} \D{\sigma} \stF{a}{\mu\nu} =0
\endEQ
for $\A{a}{\tau}$. 
Under the \YM/ gauge symmetry on $\A{a}{\mu}$, 
given by the field variation
\EQ
\sXvar\A{a}{\mu} 
= \D{\mu} \sX{a}
\endEQ
where $\sX{a}$ are arbitrary functions 
that take values in the Lie algebra $\SU{2}$,
the Lagrangian is gauge invariant, 
$\sXvar L_{\rm YM}=0$.
These gauge symmetries generate a $\SU{2}$ gauge group 
with commutator structure 
$[\sXvarsub{1}, \sXvarsub{2}] = \sXvarsub{3}$
such that $\sXsub{3}{a} = \ymLB{a}{bc} \sXsub{1}{b} \sXsub{2}{c}$.
The Lagrangian gives rise to a gauge invariant stress-energy tensor
\EQ
\T{\mu\nu}(\stF{}{}) = 
\ymCK{ab}{} \invflatmetric{\alpha\beta} (
\stF{a}{\mu\alpha} \stF{b}{\nu\beta} 
-\case{1}{4} \flatmetric{\mu\nu} \invflatmetric{\sigma\tau} 
\stF{a}{\sigma\alpha} \stF{b}{\tau\beta} )
\endEQ
which yields a causal energy-momentum for the vector potential $\A{a}{\mu}$
on spacelike hypersurfaces, 
\ie/ $\T{\mu\nu}(\stF{}{}) \t{\nu}{}$ is timelike or null 
for all unit timelike vectors $\t{\nu}{}$ on Minkowski spacetime. 
Gauge invariance of the \YM/ Lagrangian relies on the property that 
$\SU{2}$ is semisimple. 
The additional property that $\SU{2}$ is compact, 
corresponding to positive-definiteness of $\ymCK{ab}{}$, 
is essential for causality of the \YM/ stress-energy tensor 
obtained from the Lagrangian. 

Next, for formulating \FT/ theory with gauge group determined by $\G$, 
introduce as the field variable 
an antisymmetric tensor potential $\B{}{\mu\nu}$
that takes values in the Lie algebra $\G$. 
Hereafter, it is convenient to identify the vector spaces of 
$\G$ and $\SU{2}$, 
so the $\SU{2}$ basis provides a vector-space basis $\e{a}$, $a=1,2,3$,
for $\G$. 
Then the components of the antisymmetric tensor potential 
$\B{}{\mu\nu} =\B{a}{\mu\nu} \e{a}$ 
can be regarded equivalently 
as a set of three ordinary antisymmetric tensor fields $\B{a}{\mu\nu}$
on Minkowski spacetime. 
Finally, introduce the abelian field strength for $\B{a}{\mu\nu}$ 
given by the curl
\EQ\label{curlB}
\H{a}{\sigma\mu\nu} = \der{[\sigma}\B{a}{\mu\nu]} , 
\endEQ
along with its dual 
\EQ\label{dualH}
\stH{a}{\sigma} = \cross{\sigma}{\tau\mu\nu} \H{a}{\tau\mu\nu}
\endEQ
which satisfies the divergence identity
\EQ
\coder{\sigma} \stH{a}{\sigma} =0 . 
\label{FTHid}
\endEQ
Let $\ftLB{a}{bc}$ denote structure constants of $\G$,
and let $\ftTLB{ab}{c}= \ftLB{e}{bd} \ymCK{ae}{} \ymCK{}{cd}$,
where the $\SU{2}$ invariant metric provides 
a positive-definite metric $\ymCK{ab}{}$ on $\G$. 
Note this metric is not invariant 
with respect to the Lie algebra product in $\G$
unless $\G \simeq \SU{2}$. 

Now the field strength for \FT/ theory is defined in terms of 
$\B{a}{\mu\nu}$ and $\H{a}{\tau\mu\nu}$ 
by the relation 
\EQ
\K{a}{\sigma\mu\nu} + \stK{b}{[\sigma} \B{c}{\mu\nu]} \ftTLB{cb}{a}
= \H{a}{\sigma\mu\nu}
\endEQ
where
\EQ
\stK{a}{\tau} = \cross{\tau}{\sigma\mu\nu} \K{a}{\sigma\mu\nu}
\endEQ
is the dual field strength.
This field strength has a nonpolynomial expression 
in terms of $\B{c}{\mu\nu}$ 
given by 
\EQ
\stK{a}{\mu} 
= \invY{a\nu}{\mu b}(B) \stH{b}{\nu}
\endEQ
with $\invY{a\nu}{\mu b}(B)$ denoting the inverse of the tensor matrix
\EQ
\Y{a\nu}{\mu b}(B) 
= \id{\mu}{\nu} \ymCK{b}{a} 
+\cross{\mu}{\nu\sigma\tau} \ftTLB{cb}{a} \B{c}{\sigma\tau} , 
\endEQ
where $\B{c}{\sigma\tau}$ is restricted to satisfy 
$det(\Y{a\nu}{\mu b}(B))\neq 0$. 
Note the tensor matrix is symmetric
$\Y{ab}{\mu\nu}(B)=\Y{ba}{\nu\mu}(B)$
due to the antisymmetry of volume tensor $\cross{\mu\nu}{\sigma\tau}$
and the structure constants $\ftLB{a}{cb}$. 
Then, the \FT/ Lagrangian is given by 
\EQ\label{ftL}
L_{\rm FT} 
= \case{1}{2} \ymCK{ab}{} \stK{a}{\mu} \stK{b}{\nu} \Y{\mu\nu}{ab}(B) .
\endEQ
This yields the field equation for $\B{a}{\sigma\tau}$, 
\EQ
\EB{a}{\sigma\tau} = 
\cross{\sigma\tau}{\nu\mu} ( \der{\nu} \stK{a}{\mu} 
+\case{1}{2} \ftLB{a}{bc} \stK{b}{\nu} \stK{c}{\mu} ) =0 .
\endEQ
The gauge symmetry on $\B{a}{\mu\nu}$ is given by the field variation
\EQ
\vXvar\B{a}{\mu\nu} 
= \der{[\mu} \vX{a}{\nu]} -\ftTLB{cb}{a} \stK{b}{[\mu} \vX{c}{\nu]}
\endEQ
where $\vX{a}{\nu}$ are arbitrary covector functions 
that take values in the Lie algebra $G$. 
These gauge symmetries generate an abelian gauge group 
$[\vXvarsub{1}, \vXvarsub{2}] = 0$
on solutions of the field equation. 
Off solutions, the commutator structure closes to within a trivial
symmetry proportional to the field equation. 
Finally, the Lagrangian is gauge invariant to within a total divergence, 
$\vXvar L_{\rm FT} 
= \der{\mu}( \invvol{\mu\nu\sigma\tau} \frac{1}{2} \ftLB{d}{ab} \ymCK{cd}{}
\stK{a}{\sigma} \stK{b}{\tau} \vX{c}{\nu} )$.
In particular, gauge invariance holds 
without the need for $\G$ to be semisimple. 
Moreover, 
the stress-energy tensor obtained from the Lagrangian
\EQ
\T{\mu\nu}(\stK{}{}) = 
\ymCK{ab}{} (
\frac{1}{2} \stK{a}{\mu} \stK{b}{\nu} 
-\case{1}{4} \flatmetric{\mu\nu} \invflatmetric{\sigma\tau} 
\stK{a}{\sigma} \stK{b}{\tau} )
\endEQ
yields a causal energy-momentum 
for the antisymmetric tensor potential $\B{a}{\mu\nu}$
on spacelike hypersurfaces, 
\ie/ $\T{\mu\nu}(\stK{}{}) \t{\nu}{}$ is timelike or null 
for all unit timelike vectors $\t{\nu}{}$ on Minkowski spacetime.

\subsection{ Nonlinear generalization }
\label{nonlinearth}

We now construct a massless gauge theory with a nonlinear interaction
for the fields $\A{a}{\mu}$, $\B{a}{\mu\nu}$, $a=1,2,3$,
giving a novel generalization of the \YM//\FT/ theories above. 
The origin of the generalization will be explained by 
the deformation analysis carried out in \secref{deformations}.

To begin, the following algebraic structure \cite{Liealg} is needed on 
the Lie algebras $\SU{2}$ and $\G$.
Let $\auxLB{ab}{c}$ denote a bilinear map $\auxLB{}{}$ 
from $\G\times\SU{2}$ into $\SU{2}$ 
defining a representation of $\G$ on $\SU{2}$ 
\EQ
2\auxLB{[d|c}{a} \auxLB{|e]b}{c} = \auxLB{cb}{a} \ftLB{c}{de}
\label{Grepresentation}
\endEQ
such that this representation acts as a derivation 
preserving the $\SU{2}$ commutator
\EQ
\auxLB{ed}{c} \ymLB{d}{ab} = 2 \auxLB{e[a|}{d} \ymLB{c}{d|b]} .
\label{ymderivation}
\endEQ
Since $\SU{2}$ is semisimple, any derivation is given by 
an adjoint representation map
\EQ
\auxLB{eb}{c} =\ymLB{c}{db} \vsmap{e}{d}
\label{ymadrepresentation}
\endEQ
with $\vsmap{e}{d}$ denoting some linear map $\vsmap{}{}$
from $\G$ into $\SU{2}$.
Then, the relation \eqref{Grepresentation} implies that 
$\vsmap{}{}$ is a homomorphism 
(with respect to the Lie algebra product) of $\G$ into $\SU{2}$.

Consequently, if $\G$ is semisimple then 
clearly $\vsmap{}{}(\G) =\SU{2}$ and so $\G \simeq \SU{2}$
are isomorphic Lie algebras,
with the linear map $\vsmap{}{}$ being one-to-one. 
If instead $\G$ is solvable then 
the abelian two-dimensional Lie subalgebra $\U{1}{}^2$ in $\G$
is the kernel of $\vsmap{}{}$,
with $\G/\U{1}{}^2 \simeq \vsmap{}{}(\G) =\U{1}$ 
being any one-dimensional Lie subalgebra in $\SU{2}$.
Hence there are two different cases allowed for the Lie algebra structures
in the construction of the massless nonlinear theory. 
For the semisimple case when $\G \simeq \SU{2}$, 
since $\vsmap{}{}$ is an isomorphism, 
then without loss of generality it follows that 
\EQ
\vsmap{a}{b} =\kappa\id{a}{b} ,\quad
\ftLB{a}{bc} =\auxLB{bc}{a} =\kappa \ymLB{a}{bc} , 
\endEQ
where $\kappa$ is an arbitrary nonzero constant. 
Alternatively, for the solvable case when $\G=\U{1} \semi\U{1}{}^2$,
the properties of $\vsmap{}{}$ and $\G$ lead to 
\EQ
\vsmap{a}{b} =\v{}{a} \w{b}{} ,\quad
\ftLB{a}{bc} = \ftLB{a}{[b} \v{}{c]} ,\quad 
\auxLB{bc}{a} = \ymLB{a}{dc} \w{d}{} \v{}{b} 
\endEQ
for some fixed vectors $\v{a}{},\w{a}{}$ 
in the common vector space of $\G$ and $\SU{2}$,
and for some fixed linear map $\ftLB{a}{b}$ 
such that 
\EQ
\ftLB{a}{b} \v{}{a} = 0 ,\quad
\ftLB{a}{b} \v{b}{} = 0 .
\endEQ

To proceed, the construction now follows the pattern of 
the novel deformation of $\SU{2}$ \YM/ theory 
in three dimensions from \Ref{JMPpaper2}.
Let $\ftTLB{ab}{c} = \ftLB{d}{be} \ymCK{ad}{} \ymCK{}{ce}$
and $\auxTLB{a}{bc} = \auxLB{db}{e} \ymCK{}{ad} \ymCK{ce}{}$. 

Nonlinear field strengths $\P{a}{\mu\nu}$, $\Q{a}{\mu\nu\sigma}$
are introduced in terms of $\A{a}{\mu}$, $\B{a}{\mu\nu}$ by 
\EQs
&& 
\P{a}{\mu\nu} - \auxLB{bc}{a} \stQ{b}{[\mu} \A{c}{\nu]}
= \F{a}{\mu\nu} ,
\label{SUtwoP}\\
&& 
\Q{a}{\mu\nu\sigma} -\auxTLB{a}{cb} \stP{b}{[\mu\nu} \A{c}{\sigma]}
+\ftTLB{cb}{a} \stQ{b}{[\mu} \B{c}{\nu\sigma]}
= \H{a}{\mu\nu\sigma} ,
\label{SUtwoQ}
\endEQs
where 
\EQ
\stP{a}{\sigma\mu} =
\cross{\sigma\mu}{\tau\nu} \P{a}{\tau\nu} ,\quad
\stQ{a}{\sigma} = \cross{\sigma}{\tau\mu\nu} \Q{a}{\tau\mu\nu}
\label{SUtwodualPQ}
\endEQ
are the duals.
These field strengths depend nonpolynomially 
on $\A{a}{\mu}$, $\B{a}{\mu\nu}$ 
in the following form.
Define the tensor matrix
\EQ
Y(A,B)=Y\ad(A,B)= \matr{ 
\id{b}{a}\id{\mu}{\sigma}\id{\nu}{\alpha} & 
-\auxLB{bc}{a} \cross{\mu\nu}{\sigma\tau} \A{c}{\tau} \cr
-\auxTLB{a}{cb} \cross{\mu}{\sigma\alpha\tau} \A{c}{\tau} &
\id{b}{a}\id{\mu}{\sigma} 
+\ftTLB{cb}{a} \cross{\mu}{\sigma\tau\nu} \B{c}{\tau\nu} \cr 
}
\label{SUtwoY}
\endEQ
and consider the inverse matrix $Y\inv(A,B)$ satisfying
\EQ
Y\inv(A,B) Y(A,B) = Y(A,B) Y\inv(A,B) =\matr{ 
\id{b}{a}\id{\tau}{\sigma}\id{\nu}{\alpha} & 0 \cr 
0 & \id{b}{a}\id{\tau}{\sigma} } 
\endEQ
with $\A{a}{\tau}$ and $\B{a}{\tau\nu}$
restricted by the condition $\det(Y(A,B))\neq 0$ 
necessary for invertibility of $Y(A,B)$. 
Assemble the field strength duals into tensor matrices 
\EQ
N = \matr{ \stP{a}{\mu\nu} \cr \stQ{a}{\mu} } ,\quad
M = \matr{ \stF{a}{\mu\nu} \cr \stH{a}{\mu} } .
\label{SUtwoMN}
\endEQ
Then $N = Y\inv(A,B) M$, 
where $Y\inv(A,B)$ is nonpolynomial in terms of 
$\A{a}{\mu}$, $\B{a}{\mu\nu}$.

The Lagrangian for the massless nonlinear theory is constructed by 
\EQ
L_{\rm N} = 
\ymCK{ab}{} ( 
\invflatmetric{\sigma\tau} \invflatmetric{\mu\nu} 
\stP{a}{\sigma\mu} \stF{b}{\tau\nu} 
+ \invflatmetric{\sigma\tau} \stQ{a}{\sigma} \stH{b}{\tau} )
\label{SUtwoL}
\endEQ
which can be also expressed in a more symmetrical form 
$L=N\ad Y(A,B) N = M\ad Y\inv(A,B) M$. 
The gauge symmetries in this theory consist of 
the field variations given by 
\EQs
&& 
\sXvar\A{a}{\mu} 
= \D{\mu} \sX{a} + \auxLB{bc}{a} \stQ{b}{\mu} \sX{c} ,
\label{SUtwosXA}\\
&& 
\sXvar\B{a}{\mu\nu} 
= \auxTLB{a}{cb} \stP{b}{\mu\nu} \sX{c} ,
\label{SUtwosXB}
\endEQs
in terms of arbitrary functions $\sX{a}$,
and also 
\EQs
&&
\vXvar\A{a}{\mu} =0 ,
\label{SUtwovXA}\\
&&
\vXvar\B{a}{\mu\nu} 
= \der{[\mu} \vX{a}{\nu]} - \ftTLB{cb}{a} \stQ{b}{[\mu} \vX{c}{\nu]} ,
\label{SUtwovXB}
\endEQs
in terms of arbitrary covector functions $\vX{a}{\nu}$.
Under both these gauge symmetries
the Lagrangian is invariant to within a total divergence,
\EQ
\sXvar L_{\rm N} = 
\der{\mu}( 
\cross{}{\mu\nu\sigma\tau} 2 \auxLB{ac}{d} \ymCK{bd}{}
\stQ{a}{\nu} \stP{b}{\sigma\tau} \sX{c} ) ,\quad
\vXvar L_{\rm N} = 
\der{\mu}( 
\cross{}{\mu\nu\sigma\tau} \ftLB{c}{ab} \ymCK{cd}{}
\stQ{a}{\nu} \stQ{b}{\sigma} \vX{d}{\tau} ) 
\endEQ
as shown by results in \secref{deformations}.

In this construction, 
we refer to the underlying \YM//\FT/ algebraic structure $(\SU{2},\G)$
as the structure group of the massless nonlinear theory.

\subsection{ Features }

The field equations for $\A{a}{\mu}$ and $\B{a}{\mu\nu}$ 
obtained from the Lagrangian are given by 
\EQs
&& 
\EA{a}{\tau} 
= \cross{\tau}{\nu\sigma\mu} ( 
\D{\nu} \stP{a}{\sigma\mu} +\auxLB{bc}{a} \stQ{b}{\nu} \stP{c}{\sigma\mu} )
=0 ,
\label{SUtwoAeq}\\
&& 
\EB{a}{\tau\sigma}
=\cross{\tau\sigma}{\nu\mu} (
\der{\nu} \stQ{a}{\mu} 
+\case{1}{2} \ftLB{a}{bc} \stQ{b}{\nu} \stQ{c}{\mu} )
=0 .
\label{SUtwoBeq}
\endEQs
Both these field equations are of second order in derivatives of 
$\A{a}{\mu}$, $\B{a}{\mu\nu}$,
with the second derivatives appearing linearly
and first derivatives appearing quadratically,
while $\A{a}{\mu}$, $\B{a}{\mu\nu}$ appear nonpolynomially.
As a consequence of gauge invariance, 
the field equations satisfy nonlinear divergence identities
\EQs
&&
\coD{\tau} \EA{a}{\tau} 
=-\invflatmetric{\tau\mu} \auxLB{bc}{a} \stQ{b}{\mu} \EA{c}{\tau}
- \invflatmetric{\tau\mu}\invflatmetric{\sigma\nu} \auxLB{cb}{a}
\stP{b}{\mu\nu} \EB{c}{\tau\sigma} ,
\\
&&
\coder{\tau} \EB{a}{\tau\sigma} 
= -\invflatmetric{\tau\mu} \ftLB{a}{bc} \stQ{b}{\mu} \EB{c}{\tau\sigma} .
\endEQs

There are also nonlinear divergence identities that arise
on the dual field strengths 
\EQs
&& 
\invflatmetric{\sigma\mu} ( 
\D{\sigma} \stP{a}{\mu\nu} 
+ \auxLB{bc}{a} \stQ{b}{\sigma} \stP{c}{\mu\nu} )
= \invflatmetric{\sigma\mu} \auxLB{bc}{a} \EB{b}{\mu\nu} \A{c}{\sigma} ,
\label{SUtwoPid}\\
&&
\invflatmetric{\sigma\mu} ( 
\der{\sigma} \stQ{a}{\mu} 
- \ftTLB{bc}{a} \stQ{b}{\sigma} \stQ{c}{\mu} 
-\auxTLB{a}{bc} \invflatmetric{\tau\nu} \stP{b}{\sigma\tau} \stP{c}{\mu\nu} ) 
= - \invflatmetric{\sigma\mu} \auxTLB{a}{bc} \EA{c}{\mu} \A{b}{\sigma} 
- \invflatmetric{\sigma\mu} \invflatmetric{\tau\nu} \ftTLB{bc}{a}
\EB{c}{\mu\nu} \B{b}{\sigma\tau} ,
\label{SUtwoQid}
\endEQs
due to the $\SU{2}$ Bianchi identity \eqref{YMFid} on $\stF{a}{\mu\nu}$
and the linear divergence identity \eqref{FTHid} on $\stH{a}{\sigma\mu\nu}$.
Consequently, for solutions of the field equations, 
the field strengths satisfy a system of divergence and curl equations
\EQs
&& 
\D{[\nu} \stP{a}{\sigma\mu]} 
=-\auxLB{bc}{a} \stQ{b}{[\nu} \stP{c}{\sigma\mu]} ,\quad
\coD{\nu} \stP{a}{\mu\nu} 
= -\invflatmetric{\sigma\nu} \auxLB{bc}{a} \stQ{b}{\sigma} \stP{c}{\mu\nu} ,
\label{SUtwoPeq}\\
&& 
\der{[\nu} \stQ{a}{\mu]} 
= -\case{1}{2} \ftLB{a}{bc} \stQ{b}{[\nu} \stQ{c}{\mu]} ,\quad
\coder{\mu} \stQ{a}{\mu} 
=\invflatmetric{\sigma\mu} \ftTLB{bc}{a} \stQ{b}{\sigma} \stQ{c}{\mu} 
+\invflatmetric{\sigma\mu} \invflatmetric{\tau\nu} \auxTLB{a}{bc}
\stP{b}{\sigma\tau} \stP{c}{\mu\nu} ,
\label{SUtwoQeq}
\endEQs
with quadratic source terms.
In the divergence equation on $\stQ{a}{\mu}$, 
the source terms identically vanish
when $\ftLB{}{(ab)c} =\ymCK{d(a}{} \ftLB{d}{b)c}=0$, 
which occurs in the case $\G \simeq \SU{2}$. 

In both cases $\G \simeq \SU{2}$ or $\U{1} \semi\U{1}{}^2$,
the divergence and curl equations \eqrefs{SUtwoPeq}{SUtwoQeq}
together with equations \eqrefs{SUtwoP}{SUtwoQ}
constitute a first-order nonlinear field theory
for $\A{a}{\mu}$, $\B{a}{\mu\nu}$, $\stP{a}{\mu\nu}$, $\stQ{a}{\nu}$.
Moreover,
its linearization reduces to the abelian linear gauge theory of 
vector potentials and antisymmetric tensor potentials
(see \secref{linearth}),
whose field strengths represent 
free massless spin-one and spin-zero fields.
Hence, in physical terms, 
solutions of the nonlinear field strength equations 
\eqrefs{SUtwoPeq}{SUtwoQeq} 
describe a set of nonlinearly interacting
massless fields of spin-one and spin-zero, respectively,
in Minkowski spacetime. 

Under the gauge symmetries 
the field strengths have the transformation
\EQs
&&
\sXvar \stP{a}{\mu\nu} 
= \ymLB{a}{bc} \stP{b}{\mu\nu} \sX{c} 
+\tensor{(Y\inv \sX{}\cdot E)}{a}{\mu\nu} ,\quad
\sXvar \stQ{a}{\mu} 
= \tensor{(Y\inv \sX{}\cdot E)}{a}{\mu} ,
\\
&&
\vXvar \stP{a}{\mu\nu} 
= \tensor{(Y\inv \vX{}{}\cdot E)}{a}{\mu\nu} ,\quad
\vXvar \stQ{a}{\mu} 
= \tensor{(Y\inv  \vX{}{}\cdot E)}{a}{\mu} ,
\endEQs
where 
$Y\inv \sX{}\cdot E$ and $Y\inv  \vX{}{}\cdot E$
are the respective products of the inverse of the tensor matrix \eqref{SUtwoY}
with the field equation tensor matrices
\EQ
\sX{}\cdot E = 
\matr{ \auxLB{bc}{a} \EB{b}{\mu\nu} \sX{c} \cr 
-\auxTLB{a}{cb} \EA{b}{\mu} \sX{c} } ,\quad
\vX{}{}\cdot E = 
\matr{ 0 \cr 
\invflatmetric{\sigma\nu} \ftTLB{cb}{a} \EB{b}{\sigma\mu} \vX{c}{\nu} } .
\endEQ
Hence, for solutions of the field equations,
$\stP{a}{\mu\nu}$ and $\stQ{a}{\mu}$ are gauge invariant
with respect to $\vXvar$, 
while with respect to $\sXvar$, 
$\stQ{a}{\mu}$ is gauge invariant
and $\stP{a}{\mu\nu}$ transforms homogeneously 
by the adjoint representation of the Lie algebra $\SU{2}$.

The gauge symmetries on solutions of the field equations
have the commutator structure
\EQ
[\sXvarsub{1}, \sXvarsub{2}] = \sXvarsub{3} ,\quad
[\vXvarsub{1}, \vXvarsub{2}] = 0 ,\quad
[\sXvarsub{1}, \vXvarsub{1}] = 0
\endEQ
where $\sXsub{3}{a} = \ymLB{a}{bc} \sXsub{1}{b} \sXsub{2}{c}$. 
Exponentiating these gauge symmetries 
leads to a group of finite gauge transformations 
closed on solutions for $\A{a}{\mu}$, $\B{a}{\mu\nu}$. 
In particular, 
$\vXvar$ generates a $\U{1}{}^3$ 
abelian group of nonlinear gauge transformations,
while $\sXvar$ generates an $\SU{2}$ 
nonabelian group of nonlinear gauge transformations,
with $\vXvar$ and $\sXvar$ commuting.
Thus the complete gauge group for the nonlinear theory 
has the direct product structure $\SU{2} \times \U{1}{}^3$.

The spin-one field strength equations \eqref{SUtwoPeq}
lead to conserved electric and magnetic type currents 
$\J{e}{a}{\mu}=\coder{\nu}\P{a}{\mu\nu}$, 
$\J{m}{a}{\mu}=\coder{\nu}\stP{a}{\mu\nu}$
in the nonlinear theory.
Corresponding sets of electric and magnetic charges are given by 
\EQs
&& 
\charge{e}{a} = \case{1}{4\pi} \int_S 
\P{a}{\nu\mu} t^\nu dS^\mu ,\quad a=1,2,3
\label{SUtwoecharge}\\
&& 
\charge{m}{a} = \case{1}{4\pi} \int_S 
\stP{a}{\nu\mu} t^\nu dS^\mu ,\quad a=1,2,3
\label{SUtwomcharge}
\endEQs
for any closed surface $S$ in a constant time hypersurface 
in Minkowski spacetime,
with surface element $dS^\mu$ 
and hypersurface unit normal $t^\nu$. 
If the closed surface is taken to be a sphere $S_\infty$ at spatial infinity,
the resulting enclosed total charges are time-independent constants, 
$t^\nu \der{\nu} \charge{e}{a} =t^\nu \der{\nu} \charge{m}{a} =0$,
provided there is no current flow normal to $S_\infty$. 
These total charges are gauge invariant with respect to $\vXvar$
and transform by the adjoint representation of the Lie algebra $\SU{2}$
with respect to $\sXvar$
if the functions $\sX{a}$ are constant on $S_\infty$,
\EQs
&&
\sXvar\charge{e}{a} =\ymLB{a}{bc} \charge{e}{b} \sX{c} ,\quad
\sXvar\charge{m}{a} =\ymLB{a}{bc} \charge{m}{b} \sX{c} ,
\\
&&
\vXvar\charge{e}{a} =\vXvar\charge{m}{a} =0 .
\endEQs

Similarly, 
the spin-zero field strength equations \eqref{SUtwoQeq}
yield a conserved tensor
$\J{s}{a}{\sigma\mu}=\coder{\nu}\Q{a}{\sigma\mu\nu}$, 
which leads to a set of scalar type charges
\EQ
\charge{s}{a} = \case{1}{2\pi} \int_C
\Q{a}{\sigma\nu\mu} n^\sigma t^\nu ds^\mu ,\quad a=1,2,3 
\label{SUtwoscharge}
\endEQ
for any closed curve $C$ on a surface $S$ in a constant time hypersurface 
in Minkowski spacetime,
with line element $ds^\mu$, surface unit normal $n^\sigma$,
and hypersurface unit normal $t^\nu$. 
If the closed curve is taken to be a circle $C_\infty$ at spatial infinity,
the resulting enclosed total charges are time-independent constants, 
$t^\nu \der{\nu} \charge{s}{a} =0$,
provided there is no current flow normal to $C_\infty$. 
These total charges are gauge invariant with respect to 
both $\sXvar$ and $\vXvar$, 
\EQ
\sXvar\charge{s}{a} =\vXvar\charge{s}{a} =0 .
\endEQ

Note that, due to the source terms 
in the spin-one and spin-zero field strength equations,
the total charges 
\eqref{SUtwoecharge}, \eqref{SUtwomcharge}, \eqref{SUtwoscharge} 
are, in general, nonzero for solutions. 

The Lagrangian gives rise in the standard manner 
(under diffeomorphisms on Minkowski spacetime) 
to a stress-energy tensor 
\EQ
\T{\mu\nu}(\stP{}{},\stQ{}{}) = 
\ymCK{ab}{} ( 
\stP{a}{\mu\sigma} \stP{b}{\nu\tau} \invflatmetric{\sigma\tau} 
+\frac{1}{2} \stQ{a}{\mu} \stQ{b}{\nu} 
- \frac{1}{4} \flatmetric{\mu\nu} 
( \stP{a}{\sigma\alpha} \stP{b}{\tau\beta} 
\invflatmetric{\sigma\tau} \invflatmetric{\alpha\beta} 
+ \stQ{a}{\sigma} \stQ{b}{\tau} \invflatmetric{\sigma\tau} )) .
\label{SUtwoT}
\endEQ
This tensor is conserved and gauge invariant on solutions.
The conservation equation
$\coder{\mu} \T{\mu\nu}(\stP{}{},\stQ{}{}) = 0$
can be derived in a standard manner 
from the spacetime covariance of the theory,
while gauge invariance 
$\sXvar\T{\mu\nu}(\stP{}{},\stQ{}{}) = \vXvar\T{\mu\nu}(\stP{}{},\stQ{}{})=0$
manifestly holds 
due to the gauge transformation properties of the field strengths.

Conserved currents 
$\J{}{}{\mu}(\kv{}{})=\kv{\nu} \T{\mu\nu}(\stP{}{},\stQ{}{})$
are obtained from the stress-energy tensor
by contraction with a Killing vector field $\kv{\nu}$ 
on Minkowski spacetime.
These conserved currents define gauge invariant fluxes of
energy-momentum and stress carried by the fields 
on a constant time hypersurface $\Sigma$,
when $\kv{\nu}$ is taken to be a time translation and space translation,
respectively.
Fluxes of angular momentum and boost momentum are defined similarly
with $\kv{\nu}$ taken to be a rotation or boost.
In particular, for $\kv{\nu}=t^\nu$ given by 
the timelike unit normal $t^\mu$ to $\Sigma$, 
a positive energy $t^\mu t^\nu\T{\mu\nu}(\stP{}{},\stQ{}{})$ 
and a causal energy-momentum $t^\mu \T{\mu\nu}(\stP{}{},\stQ{}{})$ 
is obtained for solutions. 
The corresponding total fluxes are given by 
\EQ
\charge{}{}(\kv{}{}) 
= \int_\Sigma t^\mu \kv{\nu} \T{\mu\nu}(\stP{}{},\stQ{}{}) dV
\endEQ
where $dV$ is the volume element on $\Sigma$.

An extension of this theory from an $(\SU{2},\G)$ structure group 
to a general nonabelian structure group is presented 
in \secref{geometricalth}.

\section{ Extended deformation with Chern-Simons mass term }
\label{massiveth}

The nonlinear generalization of \YM//\FT/ gauge theories
in \secref{masslessth} has an interesting extension 
to include a \CS/ mass term. 
This construction yields a novel gauge theory 
for massive vector potentials $\A{a}{\mu}$ coupled to 
massive antisymmetric tensor potentials $\B{a}{\mu\nu}$, $a=1,2,3$,
presented here. 
For simplicity, 
the Lie algebra of the underlying \YM/ and \FT/ gauge groups
will again be given by the most general 3-dimensional possibilities,
respectively, $\SU{2}$ and $\G \simeq \SU{2}$ or $\U{1} \semi\U{1}{}^2$. 

The natural starting point is a nonabelian \CS/ type term
\cite{CSterm}
\EQ\label{csL}
L_{\rm CS} = m \invvol{\nu\mu\sigma\tau} \ymCK{ab}{}
( \B{a}{\sigma\tau} \der{\nu}\A{b}{\mu} 
+ \lambda \A{d}{\nu}\A{e}{\mu} \B{a}{\sigma\tau} \ymLB{b}{de} )
\endEQ
where $m\neq 0$ is the \CS/ mass, 
$\lambda$ is a coupling constant,
and, recall, $\ymCK{ab}{} = -\ymLB{c}{ad} \ymLB{d}{bc}$ 
is a positive definite metric on the common 3-dimensional vector space
of the Lie algebras $\SU{2},\G$.
In the case $\G=\SU{2}$, 
the addition of this Lagrangian to 
the pure \YM/ and \FT/ Lagrangians \eqrefs{ymL}{ftL} 
gives a gauge invariant Lagrangian $L=L_{\rm YM}+L_{\rm FT} +L_{\rm CS}$
if a $\SU{2}$ \YMH/ type coupling is added between
the antisymmetric tensor potentials $\B{a}{\mu\nu}$
and the \YM/ vector potentials $\A{a}{\mu}$. 
Gauge invariance also determines the \CS/ coupling to be 
$\lambda=\frac{1}{2}$. 
This yields a massive $\SU{2}$ \YM//\FT/ gauge theory \cite{FTth}
with the mass arising from the nonlinear interaction of the fields
$\A{a}{\mu}$ and $\B{a}{\mu\nu}$ through the \CS/ Lagrangian. 
The origin of the \YMH/ coupling of $\B{a}{\mu\nu}$ with $\A{a}{\mu}$
will be explained by the deformation analysis in \secref{deformations}. 
Remarkably, this coupling also allows the \CS/ Lagrangian \eqref{csL} 
to be compatible with the nonlinear generalization of
massless \YM//\FT/ theory constructed in \secref{nonlinearth},
as we now carry out. 

To begin, we replace the ordinary curl \eqref{curlB} of $\B{a}{\mu\nu}$
in the nonlinear field strengths \eqrefs{SUtwoP}{SUtwoQ}
by the \YM/ covariant curl
\EQ
\H{a}{\sigma\mu\nu} = \D{[\sigma}\B{a}{\mu\nu]}
\label{YMcurlB}
\endEQ
using the $\SU{2}$ covariant derivative operator \eqref{YMD}.
Note the dual \eqref{dualH} of $\H{a}{\sigma\mu\nu}$ 
now satisfies an $\SU{2}$ divergence identity
\EQ
\coD{\sigma} \stH{a}{\sigma} 
= \stF{b}{\mu\nu}\B{c}{\sigma\tau} \ymLB{a}{bc} 
\invflatmetric{\mu\sigma} \invflatmetric{\nu\tau} .
\label{SUtwoHid}
\endEQ
We also covariantly modify the nonlinear gauge symmetries 
\eqrefs{SUtwosXB}{SUtwovXB} on $\B{a}{\mu\nu}$
to involve an $\SU{2}$ covariant curl $\D{[\mu}\vX{a}{\nu]}$ 
in $\vXvar\B{a}{\mu\nu}$
and an $\SU{2}$ commutator $\ymLB{a}{bc} \B{b}{\mu\nu} \sX{c}$
in $\sXvar\B{a}{\mu\nu}$. 
The nonlinear gauge symmetries \eqrefs{SUtwosXA}{SUtwovXA} on $\A{a}{\mu}$
remain unchanged. 
Furthermore, 
in the algebraic structure 
used to construct the massless nonlinear theory,
the bilinear map defined by $\auxLB{ab}{c}$ 
from $\G \times \SU{2}$ into $\SU{2}$ 
remains a representation of $\G$ and a derivation of $\SU{2}$.
However, consistency of the \YMH/ coupling 
between $\B{a}{\mu\nu}$ and $\A{a}{\mu}$ requires that
the $\SU{2}$ commutator needs to act as a derivation of $\G$
\EQ
\ymLB{c}{de} \ftLB{d}{ab} = 2\ymLB{d}{[a|e} \ftLB{c}{d|b]} .
\endEQ
This holds only if $\G \simeq \SU{2}$, 
and therefore excludes the possibility $\G \simeq \U{1} \semi\U{1}{}^2$.
Hence, we thereby have 
\EQ
\ftLB{c}{ab} =\kappa \ymLB{c}{ab} ,
\endEQ
where $\kappa$ is a non-zero constant. 
Since $\auxLB{ab}{c}$ is then both 
a derivation of and representation of $\G$,
these properties fix $\auxLB{ab}{c}$ to be the adjoint representation
\EQ
\auxLB{ab}{c} = \ftLB{c}{ab} .
\endEQ

As a result, 
with the underlying \YM//\FT/ algebraic structure $\G \simeq \SU{2}$,
the nonlinear field strengths are given by 
\EQs
&& 
\P{a}{\mu\nu} - \kappa \ymLB{a}{bc} \stQ{b}{[\mu} \A{c}{\nu]}
= \F{a}{\mu\nu} ,
\label{mSUtwoP}\\
&& 
\Q{a}{\mu\nu\sigma} -\kappa \ymLB{a}{bc} ( 
\stQ{b}{[\mu} \B{c}{\nu\sigma]} - \stP{b}{[\mu\nu} \A{c}{\sigma]} )
= \H{a}{\mu\nu\sigma} ,
\label{mSUtwoQ}
\endEQs
while the nonlinear gauge symmetries take the form
\EQs
&& 
\sXvar\A{a}{\mu} 
= \D{\mu} \sX{a} + \kappa \ymLB{a}{bc} \stQ{b}{\mu} \sX{c} ,
\label{mSUtwosXA}\\
&&
\vXvar\A{a}{\mu} =0 ,
\label{mSUtwovXA}
\endEQs
and 
\EQs
&& 
\sXvar\B{a}{\mu\nu} 
= \ymLB{a}{bc} ( \B{b}{\mu\nu} + \kappa \stP{b}{\mu\nu} )\sX{c} ,
\label{mSUtwosXB}\\
&&
\vXvar\B{a}{\mu\nu} 
= \D{[\mu} \vX{a}{\nu]} + \kappa\ymLB{a}{bc} \stQ{b}{[\mu} \vX{c}{\nu]} ,
\label{mSUtwovXB}
\endEQs
in terms of arbitrary scalar functions $\sX{a}$
and covector functions $\vX{a}{\nu}$.
The complete Lagrangian is then constructed by 
adding the \CS/ Lagrangian \eqref{csL} 
to the nonlinear field strength Lagrangian \eqref{SUtwoL}, 
$L=L_{\rm N}+L_{\rm CS}$. 
This Lagrangian depends on $\A{a}{\mu}$ and $\B{a}{\mu\nu}$ 
in the nonpolynomial form
\EQ
L= M\ad Y\inv(A,B) M +m M\ad( (2-2\lambda)B +(2\lambda -1)A )
\label{mSUtwoL}
\endEQ
where $Y(A,B)$ is the symmetric tensor matrix \eqref{SUtwoY}
constructed linearly from $\A{a}{\mu},\B{a}{\mu\nu}$,
and $M$ is the tensor matrix \eqref{SUtwoMN} 
of the $\SU{2}$ field strengths of $\A{a}{\mu},\B{a}{\mu\nu}$,
and where, in the \CS/ term, 
\EQ
A=\matr{ 0 \cr \A{a}{\mu} } ,\quad 
B=\matr{ \B{a}{\mu\nu} \cr 0 }
\endEQ
are tensor matrices defined by the fields. 
Note $N= Y\inv(A,B) M$ yields 
the tensor matrix \eqref{SUtwoMN} of the nonlinear field strengths. 

Under both gauge symmetries \eqsref{mSUtwosXA}{mSUtwovXB},
the Lagrangian \eqref{mSUtwoL} is invariant to within a total divergence,
\EQs
\sXvar L = 
\der{\mu}( 
\cross{}{\mu\nu\sigma\tau} \kappa \ymCK{cd}{} \ymLB{c}{ba} 
\stQ{a}{\nu} ( 2\stP{b}{\sigma\tau} +m \B{b}{\sigma\tau} )\sX{d} ) ,
\\
\vXvar L = 
\der{\mu}( 
\cross{}{\mu\nu\sigma\tau} \ymCK{cd}{} (
\kappa \ymLB{c}{ab} \stQ{a}{\nu} \stQ{b}{\sigma} 
+m\stF{c}{\nu\sigma} )\vX{d}{\tau} ) ,
\endEQs
provided the coupling constants $\kappa$ and $\lambda$ are fixed such that
\EQ
\kappa =1/m ,\quad
\lambda = 1/2
\endEQ
as shown by results in \secref{deformations}.
This gauge theory gives a nonlinear deformation of 
the massive $\SU{2}$ \YM//\FT/ theory from \Ref{FTth}. 
We refer to the underlying algebraic structure $\SU{2}$
as the structure group of the massive nonlinear theory.

\subsection{ Features }

The Lagrangian \eqref{mSUtwoL} yields the following field equations
for $\A{a}{\mu}$ and $\B{a}{\mu\nu}$:
\EQs
&& 
\EA{a}{\tau} 
= \cross{\tau}{\nu\sigma\mu} ( 
\D{\nu} \stP{a}{\sigma\mu} 
+\ymLB{a}{bc}( \frac{1}{m} \stQ{b}{\nu} -\A{b}{\nu} ) \stP{c}{\sigma\mu} )
+m\stQ{a}{\tau}
=0 ,
\label{mSUtwoAeq}\\
&& 
\EB{a}{\tau\sigma}
=\cross{\tau\sigma}{\nu\mu} (
\D{\nu} \stQ{a}{\mu} 
+ \ymLB{a}{bc} ( \frac{1}{2m} \stQ{b}{\nu} -\A{b}{\nu} ) \stQ{c}{\mu} )
+m\stP{a}{\tau\sigma}
=0 .
\label{mSUtwoBeq}
\endEQs
These field equations are of second order in derivatives of 
$\A{a}{\mu}$, $\B{a}{\mu\nu}$,
with the second derivatives appearing linearly
and first derivatives appearing quadratically,
while $\A{a}{\mu}$, $\B{a}{\mu\nu}$ appear nonpolynomially
through the nonlinear field strengths. 
Due to the $\SU{2}$ Bianchi identity \eqref{YMFid}
and $\SU{2}$ divergence identity \eqref{SUtwoHid},
these field strengths satisfy nonlinear divergence identities 
\EQs
&& 
\invflatmetric{\nu\sigma} ( 
\D{\nu} \stP{a}{\sigma\mu} 
+\ymLB{a}{bc}( \frac{1}{m} \stQ{b}{\nu} -\A{b}{\nu} ) \stP{c}{\sigma\mu} )
= \frac{1}{m} \invflatmetric{\nu\sigma} \ymLB{a}{bc} 
\EB{b}{\nu\mu} \A{c}{\sigma} ,
\label{mSUtwoPid}\\
&&
\invflatmetric{\nu\sigma} ( 
\D{\nu} \stQ{a}{\mu} 
- \ymLB{a}{bc} \A{b}{\nu} \stQ{c}{\sigma} )
= \frac{1}{m} \invflatmetric{\nu\sigma} \ymLB{a}{bc} (
\EA{b}{\nu} \A{c}{\sigma} 
+ \invflatmetric{\mu\tau} \EB{b}{\nu\mu} \B{c}{\sigma\tau} ) .
\label{mSUtwoQid}
\endEQs
Consequently, for solutions of the field equations, 
the field strengths satisfy a system of divergence and curl equations. 
Here, in contrast to the massless nonlinear theory in \secref{nonlinearth}, 
this system can be written in terms of the field strengths alone
\EQs
&& 
\der{[\sigma} \stP{a}{\nu\mu]} +m\Q{a}{\sigma\nu\mu}
=-\frac{1}{m} \ymLB{a}{bc} \stQ{b}{[\sigma} \stP{c}{\nu\mu]} ,\quad
\coder{\nu} \stP{a}{\nu\mu} 
=-\frac{1}{m} \invflatmetric{\sigma\nu} \ymLB{a}{bc} 
\stQ{b}{\sigma} \P{c}{\nu\mu} ,
\label{mSUtwoPeq}\\
&& 
\der{[\sigma} \stQ{a}{\nu]} +m\P{a}{\sigma\nu}
= -\frac{1}{2m} \ymLB{a}{bc} \stQ{b}{[\sigma} \stQ{c}{\nu]} ,\quad
\coder{\nu} \stQ{a}{\nu} 
= 0 ,
\label{mSUtwoQeq}
\endEQs
constituting a first-order nonlinear field theory for 
$\stP{a}{\nu\mu},\stQ{a}{\nu}$. 

Linearization of the equations \eqrefs{mSUtwoPeq}{mSUtwoQeq} 
produces a system of linear massive spin-one field strength equations,
with the mass given by $m$. 
The corresponding linearization in terms of $\A{a}{\mu}$ and $\B{a}{\mu\nu}$
thus reduces to the abelian linear gauge theory of 
massive vector potentials and antisymmetric tensor potentials
(see \secref{linearth}). 
Hence, solutions of the nonlinear field theory 
for $\A{a}{\mu}$ and $\B{a}{\mu\nu}$ 
together describe a set of nonlinearly interacting massive spin-one fields
in Minkowski spacetime,
where the coupling constant of the interaction is proportional to 
the inverse mass. 
A connection between this massive nonlinear theory 
and pure massive $\SU{2}$ \YM/ theory is discussed 
in \secref{geometricalth}.

Some additional features of the massive nonlinear theory
will now be highlighted and compared to the massless nonlinear theory
from \secref{nonlinearth}.

The gauge symmetries on solutions of the field equations 
have the commutator structure 
\EQ
[\sXvarsub{1}, \sXvarsub{2}] = \sXvarsub{3} ,\quad
[\vXvarsub{1}, \vXvarsub{2}] = 0 ,\quad
[\sXvarsub{1}, \vXvarsub{1}] = \vXvarsub{3}
\endEQ
where $\sXsub{3}{a} = \ymLB{a}{bc} \sXsub{1}{b} \sXsub{2}{c}$
and $\vXsub{3}{a}{\mu} = \ymLB{a}{bc} \sXsub{1}{b} \vXsub{1}{c}{\mu}$. 
Thus the gauge group generated by exponentiation of these gauge symmetries 
is the semi-direct product $\SU{2} \semi \U{1}{}^3$,
which differs from the direct product structure 
in the massless nonlinear theory. 
Surprisingly, 
under this gauge group
the nonlinear field strengths for solutions of the field equations 
are gauge invariant 
\EQs
&&
\sXvar \stP{a}{\mu\nu} =\vXvar \stP{a}{\mu\nu} =0, \quad
\sXvar \stQ{a}{\nu} =\vXvar \stQ{a}{\nu} =0 .
\endEQs
Off solutions, the gauge symmetries are closed to within 
trivial gauge symmetries proportional to the field equations. 

Conserved electric, magnetic, and scalar type charges
are given by the same currents \eqref{SUtwoecharge}, \eqref{SUtwomcharge},
and \eqref{SUtwoscharge}
as derived for the massless nonlinear theory. 
These currents are gauge invariant on solutions of the field equations, 
due to the gauge transformation properties of the field strengths. 

More remarkably, the conserved stress-energy tensor 
obtained from the Lagrangian \eqref{mSUtwoL} 
is of the same form \eqref{SUtwoT} as in the massless nonlinear theory. 
In particular, 
the \CS/ term \eqref{csL} makes no contribution
to the stress-energy, as it has no dependence on 
the spacetime metric $\flatmetric{\mu\nu}$ 
other than through the associated (metric compatible) volume tensor
$\invvol{\mu\nu\sigma\tau}$.
This stress-energy tensor \eqref{SUtwoT} is again conserved
and gauge invariant on solutions of the field equations
in the massive nonlinear theory. 
Likewise it again yields 
a positive energy $t^\mu t^\nu\T{\mu\nu}(\stP{}{},\stQ{}{})$ 
and a causal energy-momentum $t^\mu \T{\mu\nu}(\stP{}{},\stQ{}{})$ 
carried by the fields on any constant time hyperplane,
with a unit timelike normal $t^\mu$. 

An extension of this theory from an $\SU{2}$ structure group 
to a general nonabelian structure group is presented in the next section.

\section{ Geometrical aspects }
\label{geometricalth}

The massless and massive nonlinear deformations of 
$\SU{2}$ \YM//\FT/ gauge theory 
constructed in \secrefs{masslessth}{massiveth}
have a straightforward extension from a $\SU{2}$ structure group 
to a general nonabelian structure group.
The resulting nonabelian massless and massive theories
of coupled vector and antisymmetric tensor potentials 
possess a geometrically rich structure 
involving connections on Lie group bundles 
and associated covariant derivative operators and curvatures,
which blend geometrical features of pure \YM/ theory and pure \FT/ theory,
as will be discussed here. 
In particular, this structure exposes a striking equivalence 
between the massless/massive \YM/ equations for a Lie group connection
and the field strength equations 
in the massless/massive nonlinear deformation.
An interesting duality between the massive \YM/ equations 
and massive \FT/ equations will also be noted. 

To begin, recall, the field variables consist of a set of 
three vector fields $\A{a}{\mu}$ 
and three antisymmetric tensor fields $\B{a}{\mu\nu}$, $a=1,2,3$,
with which is associated an internal three-dimensional real vector space. 
We fix a basis $\e{a}$, $a=1,2,3$, for the Lie algebra $\SU{2}$
on this vector space 
and formulate the field variables geometrically as 
an $\SU{2}$-valued 1-form 
$A=\A{a}{\mu} \e{a} d\x{\mu}$
and an $\SU{2}$-valued 2-form 
$B=\B{a}{\mu\nu} \e{a} d\x{\mu}d\x{\nu}$. 
Similarly, the nonlinear field strengths are represented geometrically as
an $\SU{2}$-valued 2-form 
$P=\P{a}{\mu\nu} \e{a} d\x{\mu}d\x{\nu}$
and an $\SU{2}$-valued 3-form 
$Q=\Q{a}{\mu\nu\sigma} \e{a} d\x{\mu}d\x{\nu}d\x{\sigma}$,
whose duals are the 2-form 
$*P=\stP{a}{\mu\nu} \e{a} d\x{\mu}d\x{\nu}$
and the 1-form 
$*Q=\stQ{a}{\mu} \e{a} d\x{\mu}$. 

We now introduce the following 
$\SU{2}$ covariant derivative operators,
using $A$ and $*Q$ as connection 1-forms:
\EQs
&& \DA{A} =d + [A,\cdot ] ,
\label{SUtwoAD}\\
&& \DA{*Q} = d +[\lambda {*Q}, \cdot ] ,
\label{SUtwoQD}\\
&& \DA{A+*Q} =d +[A+\lambda {*Q}, \cdot ] ,
\label{SUtwoAQD}
\endEQs
where $\lambda$ is a coupling constant,
and $[\cdot,\cdot]$ denotes the $\SU{2}$ Lie bracket. 
The corresponding $\SU{2}$ curvatures are given by the 2-forms
\EQs
&& \curvA{A} =dA +\frac{1}{2} [A,A] ,
\\
&& \curvA{*Q} =\lambda( d{*Q} +\frac{1}{2} \lambda [*Q,*Q] ) ,
\\
&& \curvA{A+*Q} =\curvA{A}+\curvA{*Q} +\lambda [A,*Q] ,
\endEQs
which satisfy
\EQs
(\DA{A})^2=[\curvA{A},\cdot] ,\quad
(\DA{*Q})^2=[\curvA{*Q},\cdot] ,\quad
(\DA{A+*Q})^2=[\curvA{A+*Q},\cdot ] .
\endEQs

\subsection{ Massless SU(2) Theory }
\label{geometricalSUtwoth}

The nonlinear massless field equations 
\eqrefs{SUtwoAeq}{SUtwoBeq} for $A,B$
together with the massless field strength equations 
\eqrefs{SUtwoP}{SUtwoQ} which define $P,Q$
are given in geometrical form by 
\EQs
&&
P= \curvA{A+*Q} -\curvA{*Q} ,
\\&&
Q= \DA{*Q}B +\lambda [A,*P] ,
\endEQs
and 
\EQs
&\DA{A+*Q} {*P} =0 ,&
\label{SUtwoDP}\\
&\curvA{*Q} =0 .&
\label{SUtwocurvQ}
\endEQs
Thus, $*Q$ is a zero-curvature connection,
while $*P$ is covariantly curl-free.

Hence, on solutions, 
it follows that
\EQ
P = \curvA{A+*Q}
\label{SUtwoPform}
\endEQ
is a curvature, 
while 
\EQ
Q= \DA{*Q}(B-\lambda {*P})
\label{SUtwoQform}
\endEQ
is a covariant curl. 
In addition,
the field strength identities \eqrefs{SUtwoPid}{SUtwoQid} become
\EQ
\DA{A+*Q} P =0
\endEQ
due to the $\SU{2}$ Bianchi identity, 
and 
\EQ
\DA{*Q} Q = dQ=0
\label{SUtwoDQid}
\endEQ
since $[*Q,Q]=0$ is an identity. 

Now, consider the $\SU{2}$-valued 1-form
\EQ
\ymA{SU(2)} = A+\lambda {*Q} .
\endEQ
Under the gauge symmetry $\sXvar$, 
$\ymA{SU(2)}$ transforms as a $\SU{2}$ connection
\EQ
\sXvar\ymA{SU(2)} = \ymD{\ymA{SU(2)}} \sX{}
\endEQ
where 
\EQ
\ymD{\ymA{SU(2)}} = d+[\ymA{SU(2)},\cdot] . 
\endEQ
This connection is invariant under the gauge symmetry $\vXvar$,
\EQ
\vXvar\ymA{SU(2)} = 0 . 
\endEQ
Moreover, in terms of $\ymA{SU(2)}$, 
the nonlinear field strength equations \eqrefs{SUtwoDP}{SUtwoPform} 
involving $P$ 
are simply the $\SU{2}$ \YM/ equations. 
In particular, 
\EQ
\ymF{SU(2)} = P
\endEQ
is the $\SU{2}$ curvature of $\ymA{SU(2)}$,
satisfying the \YM/ connection equation
\EQ
\ymD{\ymA{SU(2)}} *\ymF{SU(2)}=0
\endEQ
and the Bianchi identity
\EQ
\ymD{\ymA{SU(2)}} \ymF{SU(2)}=0 .
\endEQ

Similarly, consider the $\SU{2}$-valued 2-form
\EQ
\ftB{SU(2)} = B- \lambda {*P} .
\endEQ
From the field strength equation \eqref{SUtwoQform},
note $Q= \DA{*Q} \ftB{SU(2)}$ is equivalent to 
\EQ
*Q = \ftYinv{\ftB{SU(2)}}( *d\ftB{SU(2)} )
\endEQ
where $\ftYinv{\ftB{SU(2)}}$ is the inverse of the linear map 
\EQ
\ftY{\ftB{SU(2)}} = \openone +\lambda *[\ftB{SU(2)},\cdot ]
\endEQ
acting on $\SU{2}$-valued 1-forms. 
Thus, 
\EQ
\ftK{SU(2)} = {*Q}
\endEQ
is the $\SU{2}$ \FT/ 3-form field strength of $\ftB{SU(2)}$.
In particular, under the gauge symmetry $\vXvar$, 
$\ftB{SU(2)}$ transforms as a \FT/ antisymmetric tensor potential
\EQ
\vXvar\ftB{SU(2)} = \D{\ftK{SU(2)}} \vX{}{}
\endEQ
and is invariant under the gauge symmetry $\sXvar$,
\EQ
\sXvar\ftB{SU(2)} = 0 .
\endEQ
Here
\EQ
\D{\ftK{SU(2)}} = d+[*\ftK{SU(2)},\cdot]
\endEQ
is an $\SU{2}$ covariant derivative using the dual field strength 
as the connection 1-form. 
Moreover, $\ftK{SU(2)}$
satisfies both the \FT/ field equation 
\EQ
\curvA{\ftK{SU(2)}}=0
\endEQ
and field strength identity 
\EQ
d\ftK{SU(2)}=0 , 
\endEQ
which follow from 
the field strength equations \eqrefs{SUtwocurvQ}{SUtwoDQid}
involving $Q$. 

Interestingly, we therefore see that 
pure $\SU{2}$ \YM/ theory 
for a vector potential $\ymA{SU(2)}$ 
and pure $\SU{2}$ \FT/ theory 
for an antisymmetry tensor potential $\ftB{SU(2)}$
possess a combined formulation as a massless nonlinear gauge theory
given by a nonlinear deformation of 
$\SU{2}$ \YM/ gauge theory for 
$A=\ymA{SU(2)} -\lambda \ftK{SU(2)}$
and $\SU{2}$ \FT/ gauge theory for 
$B=\lambda {*\ymF{SU(2)}} + \ftB{SU(2)}$.

\subsection{ Massive SU(2) Theory }
\label{geometricalmSUtwoth}

Compared to the massless case, 
the massive nonlinear theory has some significant geometrical differences. 
The nonlinear massive field equations 
\eqrefs{mSUtwoAeq}{mSUtwoBeq} for $A,B$
together with the field strength equations 
\eqrefs{mSUtwoP}{mSUtwoQ} which define $P,Q$
take the geometrical form
\EQs
&&
P= \curvA{A+*Q} -\curvA{*Q} ,
\label{mSUPform}\\
&&
Q= \DA{A+*Q}B +\frac{1}{m} [A,*P] ,
\label{mSUtwoQform}
\endEQs
with $\lambda = 1/m$ in the covariant derivatives \eqrefs{SUtwoQD}{SUtwoAQD}, 
and 
\EQs
& \frac{1}{m} \DA{*Q} {*P} = -Q ,&
\label{mSUtwoDP}\\
& \curvA{*Q} = -P .&
\label{mSUtwocurvQ}
\endEQs
By substitution of equations \eqrefs{mSUtwocurvQ}{mSUtwoDP}
respectively into expressions \eqrefs{mSUPform}{mSUtwoQform}, 
it follows that 
\EQ
\curvA{A+*Q} =0
\label{mSUtwocurvAQ}
\endEQ
and
\EQ
\DA{A+*Q}(*P+m B) =0 .
\label{mSUtwoDPB}
\endEQ

Hence, on solutions, 
$A+\frac{1}{m} *Q$ is a zero-curvature connection,
while $*P+ m B$ is covariantly curl-free.
In addition,
the field strength identities \eqrefs{mSUtwoPid}{mSUtwoQid} become
\EQ
\DA{*Q} P =0
\label{mSUtwoDPid}
\endEQ
due to the $\SU{2}$ Bianchi identity, 
and 
\EQ
\DA{*Q} Q = dQ=0
\label{mSUtwoDQid}
\endEQ
since $[*Q,Q]=[*P,P]=0$ is an identity. 

Now, in analogy with the massless case, 
consider the $\SU{2}$-valued 1-form 
\EQ
\ymA{SU(2)} = \frac{1}{m} {*Q} .
\label{AQeq}
\endEQ
In terms of $\ymA{SU(2)}$, 
the nonlinear field strength equations 
\eqref{mSUtwoDP}, \eqref{mSUtwocurvQ}, and \eqref{mSUtwoDPid}
involving $P$ 
are simply the massive $\SU{2}$ \YM/ equations, 
in particular, 
\EQ
*\ymD{\ymA{SU(2)}} *\ymF{SU(2)} +m^2 \ymA{SU(2)}=0
\endEQ
and the Bianchi identity
\EQ
\ymD{\ymA{SU(2)}} \ymF{SU(2)}=0
\endEQ
where 
\EQ
\ymF{SU(2)} = -P
\label{FPeq}
\endEQ
is the $\SU{2}$ curvature of $\ymA{SU(2)}$.
Moreover, $\ymA{SU(2)}$ satisfies Lorentz gauge 
\EQ
d {*\ymA{SU(2)}} =0
\endEQ
due to the field strength identity \eqref{mSUtwoDQid}. 
Correspondingly, under the $\SU{2}$ gauge symmetry $\sXvar$, 
$\ymA{SU(2)}$ is gauge invariant
\EQ
\sXvar\ymA{SU(2)} = 0 .
\endEQ

Finally, 
from the remaining field strength equations \eqrefs{mSUtwocurvAQ}{mSUtwoDPB}, 
it follows that the $\SU{2}$-valued 1-form
\EQ
\ymA{flat} = A +\frac{1}{m} {*Q}
\endEQ
is a flat connection, 
with respect to which the $\SU{2}$-valued 2-form
\EQ
\ftB{curl-free} = B +\frac{1}{m} {*P}
\endEQ
is covariantly curl-free. 
Thus, up to gauge transformations, 
$\ftB{curl-free}$ is an exact 2-form
and $\ymA{flat}$ vanishes. 
This 2-form has no apparent geometrical relation to \FT/ theory,
in contrast to the situation in the massless case. 

Interestingly, however, 
the nonlinear field strength equations 
\eqref{mSUtwoDP}, \eqref{mSUtwocurvQ}, \eqref{mSUtwoDPid}, \eqref{mSUtwoDQid}
exhibit a direct relation to massive $\SU{2}$ \FT/ theory as follows. 
Consider the $\SU{2}$-valued 2-form
\EQ
\ftB{SU(2)} = -\frac{1}{m} {*P} .
\label{BPeq}
\endEQ
From the field strength equation \eqref{mSUtwoDP}, 
we see that 
\EQ
\ftK{SU(2)} = \frac{1}{m} Q
\label{KQeq}
\endEQ
is the $\SU{2}$ \FT/ field strength 3-form determined by
\EQ
\D{\ftK{SU(2)}} \ftB{SU(2)} = \ftK{SU(2)} , 
\endEQ
and hence
\EQ
\ftYinv{\ftB{SU(2)}}( *d\ftB{SU(2)} ) = *\ftK{SU(2)} .
\label{geometricalKeq}
\endEQ
We then see that the field strength equation \eqref{mSUtwocurvQ}
is simply the massive \FT/ field equation
\EQ
\curvA{\ftK{SU(2)}}= m^2 {*\ftB{SU(2)}} ,
\endEQ
while the field strength equation \eqref{mSUtwoDPid} yields
the $\SU{2}$ \FT/ field strength identity
\EQ
d\ftK{SU(2)}=0 .
\endEQ
Finally, from the field strength equation \eqref{mSUtwoDQid},
we obtain
\EQ
\D{\ftK{SU(2)}} {*\ftB{SU(2)}} =0
\endEQ
which is a nonlinear $\SU{2}$ Lorentz gauge on $\ftB{SU(2)}$.

It now follows through the duality 
\EQ
\ymA{SU(2)} = *\ftK{SU(2)} ,\quad
m\ftB{SU(2)} = *\ymF{SU(2)} ,\quad m\neq 0
\endEQ
given by equations \eqref{AQeq}, \eqref{FPeq}, \eqref{BPeq}, and \eqref{KQeq}
that this massive \FT/ theory for $\ftB{SU(2)}$
is equivalent to the massive \YM/ theory for $\ymA{SU(2)}$. 

Consequently, we see that 
pure massive $\SU{2}$ \YM/ theory 
for a vector potential $\ymA{SU(2)}$
(or equivalently pure massive $\SU{2}$ \FT/ theory 
for an antisymmetric tensor potential $\ftB{SU(2)}$), 
along with a $\SU{2}$ theory 
of a covariantly exact antisymmetric tensor potential $\ftB{curl-free}$
with respect to a flat connection $\ymA{flat}$, 
together possess a reformulation as a massive nonlinear gauge theory
given by a nonlinear deformation of 
$\SU{2}$ \YM//\FT/ theory with a \CS/ mass term 
for $A=\ymA{flat}-\ymA{SU(2)}$
and $B=\ftB{curl-free}- \frac{1}{m} {*\ymF{SU(2)}}$.

\subsection{ General Nonabelian Theory }

The $\SU{2}$ massless and massive nonlinear theories
are easily generalized so that in place of the $\SU{2}$ structure group
we have a nonabelian structure group 
based on any semisimple Lie algebra, $\vsA$.
Geometrically, $A$, $*Q$, $B$, $*P$ thereby are generalized to be
$\vsA$-valued 1-forms and 2-forms. 
The field strength equations 
in \secrefs{geometricalSUtwoth}{geometricalmSUtwoth}
retain the same geometrical form with $[\cdot,\cdot]$
given by the Lie bracket of $\vsA$. 
As shown by the deformation analysis in \secref{deformations},
this provides the most general nonabelian 
massless and massive nonlinear theories
representing a geometrical deformation of 
semisimple \YM//\FT/ gauge theory 
for Lie-algebra valued field variables $A,B$.

A further type of extension arises from considering 
non-semisimple structure groups. 
Recall from \secref{masslessth}, 
for the massless nonlinear theory
an allowed structure group is $(\SU{2},\G=\U{1} \semi\U{1}{}^2)$, 
based on using the Lie algebras 
$\SU{2}$ for the \YM/ algebraic structure
and $\G$ for the \FT/ algebraic structure
underlying the construction of the theory. 
A more general non-semisimple structure group is allowed
for both the massless as well as the massive nonlinear theories,
which will now be presented.
This extension involves some unexpected, novel algebraic features
compared to the $\SU{2}$ case. 

To proceed, we first introduce 
two Lie algebras $\vsA$ and $\vsB$
along with a homomorphism $\BintoA$ from $\vsB$ into $\vsA$.
Thus, $\BintoA(\vsB) \subseteq \vsA$ is a subalgebra of $\vsA$
while $\ker(\BintoA) \subseteq \vsB$ is an invariant subalgebra of $\vsB$. 
(In particular, note $\vsB \simeq \vsA$ are isomorphic Lie algebras
iff the kernel of $\BintoA$ is empty and $\BintoA$ is surjective.)
Then the algebraic structure 
common to both the massless and massive nonlinear theories
consists of the Lie brackets $[\cdot,\cdot]_\vsA$, $[\cdot,\cdot]_\vsB$, 
and inner products $(\cdot,\cdot)_\vsA$, $(\cdot,\cdot)_\vsB$
fixed on $\vsA$ and $\vsB$,
and a bilinear map $\bilin(\cdot,\cdot)=[\cdot,\BintoA(\cdot)]_\vsA$ 
from $\vsA \times \vsB$ into $\vsA$.
There are additional properties required to hold on this algebraic structure
in the separate massless and massive cases. 
An explanation for the origin of these properties 
will be provided by the deformation analysis in \secref{deformations}. 

We begin with some algebraic preliminaries of a techincal nature. 
Associated with the Lie brackets and bilinear map,
introduce the linear maps $\adA(\cdot)$ and $\adB(\cdot)$
denoting the adjoint representation of the Lie algebras $\vsA$ and $\vsB$, 
\EQ
\adA(v)u =[v,u]_\vsA ,\quad
\adB(v')u' =[v',u']_\vsB , 
\endEQ
and the additional linear maps $\auxAh(\cdot)$ and $\auxhA(\cdot)$
defined via $\bilin(\cdot,\cdot)$ by 
\EQs
&&
\auxAh(v')u =\adA(u)\BintoA(v')=[u,\BintoA(v')]_\vsA = \bilin(u,v') ,
\\&&
\auxhA(v)u' =\auxAh(u')v = \bilin(v,u') , 
\endEQs
for all $u,v$ in $\vsA$, and all $u',v'$ in $\vsB$. 
Let $\TBintoA(\cdot)$ denote the adjoint map of $\BintoA(\cdot)$
from $\vsA$ into $\vsB$
defined in the natural manner 
with respect to the inner products on $\vsA$ and $\vsB$. 
This gives a bilinear map $\Tbilin(\cdot,\cdot)=[\cdot,\TBintoA(\cdot)]_\vsB$ 
from $\vsB \times \vsA$ into $\vsB$.
Now, introduce the associated linear maps 
$\auxBh(\cdot)$ and $\auxhB(\cdot)$,
defined via $\Tbilin(\cdot,\cdot)$ by 
\EQs
&&
\auxBh(v)u' =\adB(u')\TBintoA(v)=[u',\TBintoA(v)]_\vsB = \Tbilin(u',v) , 
\\&&
\auxhB(v')u =\auxBh(u)v' = \Tbilin(v',u) , 
\endEQs
for all $u,v$ in $\vsA$, and all $u',v'$ in $\vsB$. 

Similarly, let $\adTA(\cdot)$ and $\adTB(\cdot)$
denote the adjoint maps of $\adA(\cdot),\adB(\cdot)$
and define the related adjoints $\adcoTA(\cdot)$ and $\adcoTB(\cdot)$ by 
\EQ
\adcoTA(v)u = \adTA(u)v ,\quad
\adcoTB(v')u' = \adTB(u')v' , 
\endEQ
as well as the analogous adjoint maps 
$\auxThA(\cdot)$, $\auxThB(\cdot)$, $\auxcoThA(\cdot)$, $\auxcoThB(\cdot)$
given via
\EQs
&&
\auxcoThA(u)v =-\TBintoA(\adcoTA(u)v) 
= -\TBintoA(\adTA(v)u) = \auxThA(v)u , 
\\&&
\auxcoThB(u')v' =-\BintoA(\adcoTB(u')v') 
= -\BintoA(\adTB(v')u') = \auxThB(v')u', 
\endEQs
again for all $u,v$ in $\vsA$, and all $u',v'$ in $\vsB$. 
Since $\BintoA$ is a homomorphism of $\vsB$ into $\vsA$,
it follows that 
\EQ\label{adrelation}
\auxcoThA(\cdot)\BintoA = -\adcoTB(\TBintoA(\cdot)) .
\endEQ

The appearance of these adjoint maps is an essential feature
in the general nonabelian algebraic structure of 
the massless and massive nonlinear theories. 
Note, we have $\adcoTA(\cdot)=\adA(\cdot)$ and $\adcoTB(\cdot)=\adB(\cdot)$ 
if and only if the the inner products are invariant 
with respect to the Lie brackets,
which holds whenever $\vsA$ and $\vsB$ are semisimple
and the inner products are given by the Cartan-Killing metrics 
of $\vsA$ and $\vsB$. 

Next, we take $A,*Q$ to be $\vsA$-,$\vsB$- valued 1-forms 
and $B,*P$ to be $\vsB$-,$\vsA$- valued 2-forms, respectively.
For later use in formulating $*Q,*P$ 
geometrically in terms of $A,B$, 
we first introduce the following 
inner product norm on pairs $(\alpha,\beta)$ consisting of 
a $\vsA$-valued 2-form $\alpha$ and a $\vsB$-valued 1-form $\beta$:
\EQ
Y_{A,B}\left( (\alpha,\beta), (\alpha,\beta) \right)
= (\alpha,\alpha)_\vsA + (\beta,\beta)_\vsB
- 2([\BintoA(\beta),A]_\vsA,*\alpha)_\vsA 
- ([\beta,\beta]_\vsB,*B)_\vsB
\label{YAB}
\endEQ
where $(\cdot,\cdot)_\vsA$ and $(\cdot,\cdot)_\vsB$ are extended to 
act on $\vsA$-,$\vsB$- valued forms via the Hodge inner product. 
Then let $Y_{A,B}(\cdot)=Y_{A,B}^{\rm T}(\cdot)$ 
be the associated symmetric linear map on pairs $(\alpha,\beta)$. 

Finally, we also introduce the following covariant derivative operators:
\EQs
&& \DA{A} =d +\adA(A) , 
\label{AD}\\
&& \DA{*Q} = d +\auxAh(*Q) , 
\label{QD}\\
&& \DA{A+*Q} =d +\adA( A+\BintoA(*Q) ) , 
\label{AQD}
\endEQs
which act on $\vsA$-valued functions and forms,
and 
\EQs
&& \DB{*Q} =d -\adTB(*Q) , 
\label{QD'}\\
&& \DB{A} =d -\auxTBhinv(A) , 
\label{AD'}\\
&& \DB{A+*Q} =d -\adTB( \BintoA\inv(A) +{*Q} ) , 
\label{AQD'}
\endEQs
which act on $\vsB$-valued functions and forms,
where the last two derivative operators are defined 
only when $\BintoA$ is invertible. 
The Lie-algebra valued curvature 2-forms 
associated with these connections are determined by
\EQs
&& \curvA{A} =dA +\frac{1}{2} [A,A]_\vsA , 
\label{AR}\\
&& \curvB{*Q} =d{*Q} +\frac{1}{2} [*Q,*Q]_\vsB , 
\label{QR'}
\endEQs
which satisfy
\EQs
&& (\DA{A})^2=\adA(\curvA{A}) , 
\\
&& (\DB{*Q})^2=-\adTB(\curvB{*Q}) .
\endEQs
Moreover, note that since $\BintoA$ is a homomorphism of $\vsB$ into $\vsA$,
we have 
\EQ
\curvA{*Q} =\BintoA(\curvB{*Q}) , 
\endEQ
while and from the property that $-\adTB(\cdot)$ is 
the coadjoint representation of $\vsB$, we also have 
\EQ
\curvB{A} =-\auxTBhinv(\curvA{A})
\endEQ
since when $\BintoA\inv$ exists 
it gives a homomorphism of $\vsA$ onto $\vsB$.

\subsubsection{ Massless Theory } 

In the massless nonlinear theory, 
the Lie algebra $\vsA$ is required to be semisimple,
and hence
\EQ
\adcoTA(\cdot)=-\adTA(\cdot)=\adA(\cdot) , 
\endEQ
with the inner product $(\cdot,\cdot)_\vsA$ given by 
the Cartan-Killing metric of $\vsA$. 
Note that, consequently, 
\EQ
\auxTBh(\cdot) =\TBintoA\circ\auxAh(\cdot) 
\endEQ
and so
\EQ
\TBintoA( \DA{*Q}(\cdot) ) = \DB{*Q}( \TBintoA(\cdot) ) . 
\endEQ

No further properties are needed on the Lie algebra structure of $\vsA,\vsB$.
Now the entire theory can be constructed geometrically
in terms of the covariant derivatives \eqref{AD}, \eqref{AQD}, \eqref{QD'}
and curvatures \eqrefs{AR}{QR'}
along with the linear map \eqref{YAB}. 
First,
the massless nonlinear field strengths are defined by
\EQ\label{masslessPQ}
(*P,*Q) = Y_{A,B}^{-1}(*\curvA{A},*dB)
\endEQ
where $Y_{A,B}^{-1}(\cdot)$ is the inverse of the linear map $Y_{A,B}(\cdot)$. 
In terms of these field strengths,
the gauge symmetries on $A,B$ are given by
\EQ
\sXvar A = \DA{A+*Q}\xi ,\quad
\sXvar B = \conxB{*P}\xi , 
\endEQ
for arbitrary $\vsA$-valued functions $\xi$ on $M$,
and 
\EQ
\vXvar A = 0 ,\quad
\vXvar B = \DB{*Q}\chi , 
\endEQ
for arbitrary $\vsB$-valued 1-forms $\chi$ on $M$. 
Here $\conxB{*P}(\cdot)$ is a linear map 
associated with $*P$ by 
\EQ
\conxB{*P} = \auxThB(*P) . 
\endEQ
Finally, the Lagrangian is given by 
\EQ
L= \frac{1}{2} (*P,\curvA{A})_\vsA + \frac{1}{2} (*Q,*dB)_\vsB
= \frac{1}{2} (*P,*Q) \cdot Y_{A,B}(*P,*Q) 
\endEQ
where $(\alpha,\beta)\cdot(\alpha,\beta) 
= (\alpha,\alpha)_\vsA + (\beta,\beta)_\vsB$
for any $\vsA$-valued forms $\alpha$, $\vsB$-valued forms $\beta$. 
This yields the field equations for $A,B$:
\EQ
*\EA{}{} = \DA{A+*Q} {*P} =0 ,\quad
*\EB{}{} = \curvB{*Q} =0 . 
\endEQ
Thus, on solutions, $*Q$ is a zero-curvature connection,
while $*P$ is covariantly curl-free.

From the field strength equation \eqref{masslessPQ}, 
$*P$ and $*Q$ have the form 
\EQ
P= \curvA{A+*Q} -\curvA{*Q} ,\quad
Q= \DB{*Q}B -\conxB{*P}A . 
\endEQ
Hence, since $\curvA{*Q}=\BintoA(\curvB{*Q}) =0$ 
and $\adA(*P)A =\DA{*Q} {*P}$ on solutions, 
it respectively follows that
\EQ
P = \curvA{A+*Q}
\endEQ
is a curvature, 
while 
\EQ
Q= \DB{*Q}( B-\TBintoA(*P) )
\endEQ
is a covariant curl,
using in addition the algebraic relation
\EQ
\auxThA(\cdot) = \TBintoA\circ\adA(\cdot) . 
\endEQ
Then, 
the covariant exterior derivatives 
$\DB{*Q}$ of $Q$ and $\DA{A+*Q}$ of $P$ yield
\EQ
\DA{A+*Q} P =0
\endEQ
and 
\EQ
\DB{*Q} Q =0 . 
\endEQ
These are the same geometrical expressions as those in the $\SU{2}$ case. 

Therefore, geometrically, $\adA(A+\TBintoA(*Q))=\ymA{YM}$ 
is a \YM/ connection 1-form,
whose curvature $\adA(P)=\ymF{YM}$
satisfies the massless \YM/ equations $\ymD{\ymA{YM}} *\ymF{YM}=0$
and the Bianchi identity $\ymD{\ymA{YM}} \ymF{YM}=0$,
with $\ymD{\ymA{YM}}=d+\adA(\ymA{YM})$, 
based on the gauge group associated to the semisimple Lie algebra $\vsA$. 

\subsubsection{ Massive Theory }

In the massive nonlinear theory, 
the homomorphism $\BintoA$ is required to be 
a Lie-algebra isomorphism
\EQ
\BintoA(\cdot) = \frac{1}{m} \idmap(\cdot)
\endEQ
so $\vsA =\BintoA(\vsB) \simeq \vsB$, 
where $m\neq 0$ is the mass,
and $\idmap$ is a linear map 
identifying the vector spaces of $\vsA$ and $\vsB$.
But, $\vsA$ and $\vsB$ need not be semisimple here,
and there are no further properties required on 
the Lie algebra structure of $\vsA,\vsB$.
Thus, surprisingly, 
compared to massive \YM//\FT/ theory \cite{FTth}
as well as to pure massless \YM/ theory, 
a more general structure group is allowed
for the massive nonlinear theory. 

First,
the massive nonlinear field strengths are defined by
\EQ\label{massivePQ}
(*P,*Q) = Y_{A,B}^{-1}(*\curvA{A},*\DB{A}B)
\endEQ
where $Y_{A,B}^{-1}(\cdot)$ is the inverse of the linear map $Y_{A,B}(\cdot)$. 
In terms of these field strengths,
the gauge symmetries on $A,B$ are given by
\EQ
\sXvar A = \DA{A+*Q}\xi ,\quad
\sXvar B = \conxB{B+*P}\xi , 
\endEQ
for arbitrary $\vsA$-valued functions $\xi$ on $M$,
and 
\EQ
\vXvar A = 0 ,\quad
\vXvar B = \DB{A+*Q}\chi , 
\endEQ
for arbitrary $\vsB$-valued 1-forms $\chi$ on $M$,
where now 
\EQ
\conxB{B+*P} = -\auxcoThB(*P) +\auxhB(B) . 
\endEQ
The Lagrangian is given by 
\EQ
L= \frac{1}{2} (*P+m^2\BintoA(B),\curvA{A})_\vsA 
+ \frac{1}{2} (*Q,*\DB{A}B)_\vsB , 
\endEQ
which yields the field equations for $A,B$:
\EQ
*\EA{}{} = \BintoA( \DB{*Q} \BintoA\inv(*P) ) +m^2\BintoA(Q) =0 ,\quad
*\EB{}{} = \curvB{*Q} +\BintoA\inv(P) =0 
\endEQ
where $\BintoA\inv$ is the inverse of the isomorphism $\BintoA$,
\EQ
\BintoA\inv(\cdot) = m \idmap(\cdot) = m^2\TBintoA(\cdot) . 
\endEQ

From the field strength equation \eqref{massivePQ}, 
$*P$ and $*Q$ have the form 
\EQ
P= \curvA{A+*Q} -\curvA{*Q} ,\quad
Q= \DB{A+*Q}B -\conxB{*P}A . 
\endEQ
Hence, 
since 
\EQ
\curvA{*Q} =-P
\label{curvQPeq}
\endEQ 
holds on solutions, 
it follows that
\EQ
\curvA{A+*Q} =0
\endEQ
and so $A+*Q$ is a zero-curvature connection. 
Furthermore, from 
\EQ
\DB{*Q} \TBintoA(*P) =-Q
\label{curldualPQeq}
\endEQ
on solutions, 
and from $\conxB{*P}A = \auxTBhinv(A)\TBintoA(*P)$
through the algebraic relation
\EQ
\auxThA(\cdot) = -\auxTBhinv(\cdot)\TBintoA
\endEQ
obtained from the homomorphism equation \eqref{adrelation}, 
it follows that 
\EQ
\DB{A+*Q}(B+\TBintoA(*P)) =0
\endEQ
and so $B+\TBintoA(*P)$ is covariantly curl-free.
Then, 
the covariant exterior derivatives 
$\DB{*Q}$ of $Q$ and $\DA{*Q}$ of $P$ yield
\EQ
\DA{*Q} P =0
\label{curlPQeq}
\endEQ
and 
\EQ
\DB{*Q} Q = \TBintoA( \adTA(*P) P ) . 
\label{curlQQeq}
\endEQ

Therefore, geometrically, $\auxAh(*Q)=\ymA{YM}$ 
is a \YM/ connection 1-form,
whose curvature $-\adA(P)=\ymF{YM}$
satisfies an adjoint version of the massive \YM/ equations 
\EQ
*\ymTD{\ymA{YM}} *\ymF{YM}+m^2 \ymA{YM}=0 
\label{adjYMeq}
\endEQ
and the Bianchi identity 
\EQ
\ymTD{\ymA{YM}} \ymF{YM}=0 ,
\endEQ
where 
\EQ
\ymTD{\ymA{YM}} = d-\adTA(\ymA{YM}) .
\endEQ
In addition, $\ymA{YM}$ satisfies a nonlinear covariant gauge condition
\EQ
\ymTD{\ymA{YM}} *\ymA{YM} = -\adTA(*\ymF{YM})\ymF{YM} .
\label{adjYMgauge}
\endEQ
Interestingly, 
this adjoint modification is based on having a non-semisimple 
Lie algebra $\vsA$, so that $\adcoTA(\cdot) \neq \adA(\cdot)$. 
Its consistency relies on the property that, for any Lie algebra $\vsA$,
$-\adTA(\cdot)$ is the coadjoint representation of $\vsA$. 
If $\vsA$ is chosen to be semisimple, 
then note the standard massive \YM/ theory is obtained. 

Similarly to the $\SU{2}$ case, 
the non-semisimple massive \YM/ theory 
\eqsref{adjYMeq}{adjYMgauge}
here is equivalent to a non-semisimple massive \FT/ theory 
given by the duality 
\EQ
\ymA{YM} = *\ftK{YM} ,\quad
m\ftB{YM} = *\ymF{YM} 
\endEQ
as follows from the field strength equations
\eqref{curvQPeq}, \eqref{curldualPQeq}, 
\eqref{curlPQeq} and \eqref{curlQQeq}.

\section{ Deformation analysis }
\label{deformations}

Here a systematic determination of 
the most general nonlinear geometrical deformation will be given for
the linear gauge theory of 
$n\ge 1$ vector potentials $\A{a}{\mu}$, 
$a=1,\ldots,n$, 
and $n'\ge 1$ antisymmetric tensor potentials $\B{a'}{\mu\nu}$, 
$a'=1,\ldots,n'$,
with a \CS/ type mass term, 
on a 4-dimensional spacetime manifold $M$. 
The method used is a geometrical version of the field theoretic approach
to deformations developed in \Ref{AMSpaper,JMPpaper1,Annalspaper}.

\subsection{ Linear theory }
\label{linearth}

We formulate the linear theory geometrically, 
using a set of 
1-forms $\A{a}{} =\A{a}{\mu} d\x{\mu}$
and 2-forms $\B{a'}{}=\B{a'}{\mu\nu} d\x{\mu}d\x{\nu}$.
These field variables are regarded as taking values 
in respective internal vector spaces $\vsA,\vsB$ of dimensions $n,n'$. 

To proceed, the only structure we require on the spacetime manifold $M$ is 
the exterior derivative operator $d$
and the Hodge dual $*$ such that $*^2=\pm \openone$
where $\openone$ is the identity operator. 
Hereafter, products of fields will be understood to be 
wedge products of forms on $M$ 
(and tensor products with respect to $\vsA,\vsB$). 
Recall, in terms of $*$, there is a standard Hodge inner product 
on pairs $(\alpha,\beta)$ of 1-forms and 2-forms, 
$(\alpha,\beta)\cdot(\alpha,\beta) 
= *(\alpha *\alpha) - *(\beta *\beta)$. 

The linear field strengths associated with the field variables 
are given by 
the $\vsA$-valued 2-form $\linF{a}=d\A{a}{}$
and $\vsB$-valued 3-form $\linH{a'} =d\B{a'}{}$. 
Then the Lagrangian is given by the following real-valued 4-form 
\EQ
\nthL{2} =
\frac{1}{2}\id{ab}{} \linF{a} *\linF{b} 
- \frac{1}{2}\id{a'b'}{} \linH{a'} *\linH{b'}
+ \m{aa'}{} \linF{a} \B{b'}{}
\label{quadL}
\endEQ
where $\id{ab}{},\id{a'b'}{}$ represent components of 
respective inner products on $\vsA,\vsB$, 
and $\m{aa'}{}$ represents components of a bilinear form 
on $\vsA\times\vsB$. 
We refer to $\m{aa'}{}$ as the mass tensor. 
This Lagrangian is invariant to within an exact 4-form 
under the separate abelian gauge symmetries given by 
\EQ
\nthsXvar{0}\A{a}{} =d\sX{a} ,\quad
\nthsXvar{0}\B{a'}{} = 0 , 
\label{zerosXAB}
\endEQ
for arbitrary $\vsA$-valued functions $\sX{a}$,
and 
\EQ
\nthvXvar{0}\A{a}{} = 0 ,\quad
\nthsXvar{0}\B{a'}{} = d\vX{a'}{} ,\quad
\label{zerovXAB}
\endEQ
for arbitrary $\vsB$-valued 1-forms $\vX{a'}{}$. 
Under variations of the fields $\A{a}{}$ and $\B{a'}{}$, 
the Lagrangian yields the Euler-Lagrange field equations
\EQs
&&
*\nthEA{1}{a} = d {*\linF{a}} + \m{a'}{a} \linH{a'} =0 , 
\label{linEA}\\
&&
*\nthEB{1}{a'} = d {*\linH{a'}} + \Tm{a}{a'} \linF{a} =0 , 
\label{linEB}
\endEQs
where $\m{a'}{a}=\id{}{ab}\m{ba'}{}$ and $\Tm{a}{a'}=\id{}{a'b'}\m{ab'}{}$
are components of linear maps 
$\mA(\cdot)$ from $\vsA$ into $\vsB$, 
and $\mB(\cdot)$ from $\vsB$ into $\vsA$. 

Note, in the case when the mass tensor vanishes, 
$\m{aa'}{}=0$, 
the fields $\A{a}{}$ and $\B{a'}{}$ are decoupled
and the linear theory reduces to 
massless abelian \YM/ gauge theory for $\A{a}{}$
and massless abelian \FT/ gauge theory for $\B{a'}{}$.
The field strengths $\linF{a}$ and $\linH{a'}$
obviously then describe free massless spin-one and spin-zero fields. 

In contrast, in the opposite case when the mass tensor is fully nondegenerate, 
$\m{aa'}{}=m\id{aa'}{}$ 
where $\id{a}{a'}$ is a vector-space isomorphism of $\vsA$ and $\vsB$
(and hence $n=n'$), 
the fields $\A{a}{}$ and $\B{a'}{}$ are coupled through a \CS/ mass term. 
The linear theory then reduces to massive abelian \YM//\FT/ gauge theory,
which is the linearization of the nonlinear theory given in \Ref{FTth}. 
Consequently, the field strengths $\linF{a}$ and $\linH{a'}$
together describe free massive spin-one fields,
with the mass given by $m$. 
(In particular, 
$\A{a}{}$ supplies two of the three spin-one helicity components 
while $\B{a'}{}$ supplies the third.)

To continue, we consider the general case
with no conditions assumed on the mass tensor. 
Let $\vsA_0$ and $\vsB_0$ denote the kernels of the maps 
$\mA(\cdot)$ and $\mB(\cdot)$, 
and let $\vsA_{\rm m}$ and $\vsB_{\rm m}$ 
denote the orthogonal complements of these kernels. 
Note there is a direct sum decomposition, 
$\vsA = \vsA_0 \oplus \vsA_{\rm m}$, $\vsB=\vsB_0 \oplus \vsB_{\rm m}$,
with respect to the inner products
on the internal vector spaces. 
Moreover, $\vsA_{\rm m}$ and $\vsB_{\rm m}$ are isomorphic vector subspaces,
with a common dimension denoted by $0\le k\le n,n'$. 

Fix a basis for these vector subspaces so that 
the fields $\A{a}{}$ and $\B{a'}{}$ 
belong to $\vsA_0$ and $\vsB_0$ for $a=a'=1,\ldots,k$ 
and belong to $\vsA_{\rm m}$ and $\vsB_{\rm m}$ 
for $a=k+1,\ldots,n$, $a'=k+1,\ldots,n'$.
Then, physically speaking, 
it follows from the linear field equations that 
the field strengths $\linF{a}$ and $\linH{a'}$ given by $a=a'=1,\ldots,k$, 
together describe a set of $k$ free massive spin-one fields 
with mass equal to the non-zero eigenvalues of $\m{aa'}{}$, 
while the remaining field strengths $\linF{a}$ and $\linH{a'}$ 
given by $a=k+1,\ldots,n$ and $a'=k+1,\ldots,n'$
describe separate sets of $n-k$ free massless spin-one fields 
and $n'-k$ free massless spin-zero fields, respectively.

\subsection{ Determining equations for nonlinear deformations }
\label{deteqs}

We now consider nonlinear deformations of the linear abelian gauge theory
for $\A{a}{}$ and $\B{a'}{}$, 
with the deformation terms being locally constructed 
in a geometrical manner from the fields 
by using only the exterior derivative $d$ and Hodge dual $*$ on $M$. 
Here, a deformation consists of 
adding linear and higher power terms 
to the abelian gauge symmetries \eqrefs{zerosXAB}{zerovXAB},
\EQ
\sXvar\A{a}{} =\nthsXvar{0}\A{a}{} + \nthsXvar{1}\A{a}{} +\cdots ,\quad
\sXvar\B{a'}{} = \nthsXvar{0}\B{a'}{} +\nthsXvar{1}\B{a'}{} +\cdots , 
\label{deformsXvar}
\endEQ
and 
\EQ
\vXvar\A{a}{} =\nthvXvar{0}\A{a}{} + \nthvXvar{1}\A{a}{} +\cdots ,\quad
\vXvar\B{a'}{} = \nthvXvar{0}\B{a'}{} +\nthvXvar{1}\B{a'}{} +\cdots , 
\label{deformvXvar}
\endEQ
while simultaneously adding quadratic and higher power terms to 
the linear field equations \eqrefs{linEA}{linEB},
\EQs
&&
\EA{a}{} = \nthEA{1}{a} + \nthEA{2}{a} +\cdots ,\quad
\EB{a'}{} = \nthEB{1}{a'} + \nthEB{2}{a'} +\cdots , 
\label{deformEAB}
\endEQs
such that there exists a locally constructed Lagrangian 4-form
that is gauge invariant to within an exact 4-form. 
The condition of gauge invariance is expressed by 
\EQs
&&
\sXvar L = 
\nthsXvar{0}\nthL{2} + \nthsXvar{1}\nthL{2} + \nthsXvar{0}\nthL{3} + \cdots
=d\striv , 
\\
&&
\vXvar L = 
\nthvXvar{0}\nthL{2} + \nthvXvar{1}\nthL{2} + \nthvXvar{0}\nthL{3} + \cdots
=d\vtriv , 
\endEQs
holding for some locally constructed 3-forms $\striv$ and $\vtriv$,
where the Lagrangian is related to the field equations through
\EQ
\delta L = 
\delta\A{a}{} *\EA{b}{} \id{ab}{} 
+ \delta\B{a'}{} *\EB{b'}{} \id{a'b'}{} 
+d\Gamma , 
\endEQ
holding for some locally constructed 3-form $\Gamma$, 
under arbitrary variations $\delta\A{a}{},\delta\B{a'}{}$.

For writing down deformation terms and analyzing the deformation equations,
a precise formal setting is provided by 
the field space $\fieldsp=\{ (\A{a}{}(x),\B{a'}{}(x)) \}$
defined as the set of all sections of the vector bundle of 
$\vsA$-valued 1-forms and $\vsB$-valued 2-forms on $M$. 
Hereafter, geometrically, 
a field variation $(\delta\A{a}{},\delta\B{a'}{})$ 
is regarded as a vector field on $\fieldsp$
while field equations $(\coEA{a}{},\coEB{a'}{})$
are regarded as a covector field on $\fieldsp$,
where $\coEA{a}{},\coEB{a'}{}$ are related to $\EA{a}{},\EB{a'}{}$ by 
\EQ
\id{ab}{} *\EA{a}{} \cdot \delta\A{b}{} = 
\delta\A{a}{} \intprod \coEA{a}{} ,\quad
\id{a'b'}{} *\EA{a'}{} \cdot \delta\B{b'}{} = 
\delta\B{a'}{} \intprod \coEB{a'}{} . 
\endEQ
Here the hook $\intprod$ denotes interior product of 
a vector field with a covector field on $\fieldsp$. 
Associated to $\fieldsp$ is the jet space defined using local coordinates
\EQ
\jetsp= 
(x,\A{a}{},\B{a'}{},d\A{a}{},d\B{a'}{},d{*d\A{a}{}},d{*d\B{a'}{}},\ldots)
\endEQ
where $x$ represents a point in $M$; 
$\A{a}{}$, $d\A{a}{}$, $d{*d\A{a}{}}$, $\ldots$
represent the values of the $\vsA$-valued 1-form field $\A{a}{}(x)$
and its exterior derivatives at $x$; 
and $\B{a'}{}$, $d\B{a'}{}$, $d{*d\B{a'}{}}$, $\ldots$
represent the values of the $\vsB$-valued 2-form field $\B{a'}{}(x)$
and its exterior derivatives at $x$. 
In this setting, a locally constructed function or $p$-form on $M$
is a function purely of the jet variables 
$(\A{a}{},\B{a'}{},d\A{a}{},d\B{a'}{},d{*d\A{a}{}},d{*d\B{a'}{}},\ldots)$
and their Hodge duals 
$(*\A{a}{},*\B{a'}{},*d\A{a}{},*d\B{a'}{},
*d{*d\A{a}{}},*d{*d\B{a'}{}},\ldots)$, 
up to some finite order. 
Let $\Ader{a}$, $\Bder{a'}$, and $\kthdAder{a}$, $\kthdBder{a'}$, 
$k=0,1,2,\ldots$, 
denote derivatives with respect to the jet variables. 
Note the derivatives $\Ader{a},\Bder{a'}$ 
produce covector fields on $\fieldsp$.
We define contravariant derivatives $\Acoder{a},\Bcoder{a'}$ 
that produce vector fields on $\fieldsp$ via the natural pairing
$\Acoder{a} g \intprod \Ader{a} f = 
\id{}{ab} \Ader{a} g \cdot \Ader{b} f$
and 
$\Bcoder{a'} g \intprod \Bder{a'} f
=\id{}{a'b'} \Bcoder{a'} g \cdot\Bder{b'} f$, 
for any locally constructed functions $f,g$. 
Likewise we define $\kthdAcoder{a},\kthdBcoder{a'}$. 
Then, we introduce \EL/ operators given by 
\EQ
\ELop{\A{a}{}}= \Acoder{a} - \sum_{k\ge 0} (*d*)^{k+1} \kthdAcoder{a} ,\quad
\ELop{\B{a'}{}}= \Bcoder{a'} - \sum_{k\ge 0} (*d*)^{k+1} \kthdBcoder{a'} .
\endEQ
These operators take locally constructed functions $f$ 
into vector fields $(\ELop{\A{a}{}}(f),\ELop{\B{a'}{}}(f))$ on $\fieldsp$
and have the property that 
$\ELop{\A{a}{}}(f)=\ELop{\B{a'}{}}(f)=0$
annihilates a locally constructed function $f$ 
if and only if $*f=d\Gamma$ 
for some locally constructed 3-form $\Gamma$. 
The related operators 
\EQ
\coELop{\A{a}{}}
= \Ader{a} - \sum_{k\ge 0} (*d*)^{\sharp k+1} \kthdAder{a} ,\quad
\coELop{\B{a'}{}}
= \Bder{a'} - \sum_{k\ge 0} (*d*)^{\sharp k+1} \kthdBder{a'} 
\endEQ
yield covector fields on $\fieldsp$,
where $*^\sharp$ and $d^\sharp$ denote 
the contravariant Hodge dual operator
and contravariant exterior derivative operator
on vectors and antisymmetric tensors on $M$. 

The relation between the deformation terms 
in the field equations and Lagrangian 
is most naturally expressed through the \EL/ operators by
\EQ
\nthcoEA{k}{a} = \coELop{\A{a}{}}(*\nthL{k+1}) ,\quad
\nthcoEB{k}{a'} = \coELop{\B{a'}{}}(*\nthL{k+1})  , 
\endEQ
which determines
\EQ
\nthL{k+1} = -
\frac{1}{k+1} ( \nthEA{k}{a} \A{b}{}\id{ab}{} 
- \nthEB{k}{a'} \B{b'}{}\id{a'b'}{} )
\endEQ
to within an exact 4-form. 

In terms of the \EL/ operators, 
the condition for existence of a gauge-invariant Lagrangian 
is equivalent to the equations
\EQs
&& 
\ELop{\A{a}{}}( 
\sXvar\A{a}{}\cdot *\EA{b}{} \id{ab}{} 
+ \sXvar\B{a'}{}\cdot *\EB{b'}{} \id{a'b'}{} ) 
=0 , 
\\
&& 
\ELop{\A{a}{}}( 
\vXvar\A{a}{}\cdot *\EA{b}{} \id{ab}{} 
+ \vXvar\B{a'}{}\cdot *\EB{b'}{} \id{a'b'}{} ) 
=0 , 
\\
&& 
\ELop{\B{a'}{}}( 
\sXvar\A{a}{}\cdot *\EA{b}{} \id{ab}{} 
+ \sXvar\B{a'}{}\cdot *\EB{b'}{} \id{a'b'}{} ) 
=0 , 
\\
&&
\ELop{\B{a'}{}}( 
\vXvar\A{a}{}\cdot *\EA{b}{} \id{ab}{} 
+ \vXvar\B{a'}{}\cdot *\EB{b'}{} \id{a'b'}{} ) 
=0 .
\endEQs
These four equations are the determining system for all allowed deformations. 

\Proclaim{ Remark }{
To proceed, we restrict attention to deformations that involve 
at most one derivative of $\A{a}{}$, $\B{a'}{}$, $\sX{a}$, $\vX{a'}{}$
in the gauge symmetries 
and at most two derivatives of $\A{a}{}$, $\B{a'}{}$
in the field equations. 
Such deformations automatically preserve the number of 
gauge degrees of freedom and initial-data degrees of freedom 
in the nonlinear theory to be the same as those in the linear theory. 
Also, we consider only nontrivial deformations such that
the field equations and gauge symmetries in the nonlinear theory 
are not equivalent to those in the linear theory 
by a change either of field variables 
or of gauge symmetry variables 
(see \cite{AMSpaper}). 
}

The determining system can be reformulated more usefully and geometrically
as Lie derivative equations. 
We introduce the Lie derivative $\Lie{\delta}$ 
with respect to a vector field $(\delta\A{a}{},\delta\B{a'}{})$
on $\fieldsp$ 
acting on a locally constructed covector field 
$(\covec{a}{A},\covec{a'}{B})$ by
\EQs
(\Lie{\delta} f)\mixedindices{\rm A}{a} = && 
( \delta\A{b}{} \intprod \Ader{b}\covec{a}{A}
+ \delta\B{b'}{} \intprod \Bder{b'}\covec{a}{A}
+ \sum_{k\ge 0} (d*)^k d\delta\A{b}{} \intprod \kthdAder{b}\covec{a}{A}
+ (d*)^k d\delta\B{b'}{} \intprod \kthdBder{b'}\covec{a}{A} )
\nonumber\\&&
-( \Ader{a}\delta\A{b}{} \intprod \covec{b}{A} 
+  \Ader{a}\delta\B{b'}{} \intprod \covec{b'}{B} 
- \sum_{k\ge 0} (*d*)^{\sharp k+1} (
\kthdAder{a}\delta\A{b}{} \intprod \covec{b}{A} 
+ \kthdAder{a}\delta\B{b'}{} \intprod \covec{b'}{B} ) ) , 
\nonumber\\
\\
(\Lie{\delta} f)\mixedindices{\rm B}{a'} = &&
( \delta\A{b}{} \intprod \Ader{b}\covec{a'}{B}
+ \delta\B{b'}{} \intprod \Bder{b'}\covec{a'}{B}
+ \sum_{k\ge 0} (d*)^k d\delta\A{b}{} \intprod \kthdAder{b}\covec{a'}{B}
+ (d*)^k d\delta\B{b'}{} \intprod \kthdBder{b'}\covec{a'}{B} )
\nonumber\\&&
-( \Bder{a'}\delta\A{b}{} \intprod \covec{b}{A} 
+ \Bder{a'}\delta\B{b'}{} \intprod \covec{b'}{B} 
- \sum_{k\ge 0} (*d*)^{\sharp k+1} (
\dBder{a'}\delta\A{b}{} \intprod \covec{b}{A} 
+ \dBder{a'}\delta\B{b'}{} \intprod \covec{b'}{B} ) ) .
\nonumber\\
\endEQs

\Proclaim{ Theorem~1 }{
Local gauge invariance holds if and only if 
the Lie derivative of the field equations
with respect to the gauge symmetries vanishes
\EQ
\Lie{\sXvar}(\coEA{a}{},\coEB{a'}{}) = 0 ,\quad
\Lie{\vXvar}(\coEA{a}{},\coEB{a'}{}) = 0 .
\endEQ
}

Geometrically, these equations assert that the gauge symmetries
are vector fields tangential to the surface in $\fieldsp$ 
corresponding to the field equations.
Due to gauge invariance, 
the commutators of these vector fields have the same property.

\Proclaim{ Theorem~2 }{
Local gauge invariance holds only if 
the Lie derivative of the field equations
with respect to the gauge symmetry commutators vanishes
\EQ
\Lie{[\sXvarsub{1},\sXvarsub{2}]}(\coEA{a}{},\coEB{a'}{}) = 0 ,\quad
\Lie{[\vXvarsub{1},\vXvarsub{2}]}(\coEA{a}{},\coEB{a'}{}) = 0 ,\quad
\Lie{[\sXvarsub{1},\vXvarsub{1}]}(\coEA{a}{},\coEB{a'}{}) = 0 .
\endEQ
}

An expansion of these equations in powers of $\A{a}{}$ and $\B{a'}{}$
(and their derivatives) gives a hierarchy of determining equations
whose solutions yield all allowed deformation terms
in the field equations and gauge symmetries. 
We now find the solution of these determining equations 
explicitly at the lowest orders to give all first-order deformations 
and then outline an induction analysis to obtain a uniqueness result 
for the higher-order deformations.

\subsection{ First-order deformations 
and uniqueness of higher-order deformations }

Up to a change of field variables and gauge symmetry variables,
the most general possible first-order deformation terms 
for the gauge symmetries are given by 
\EQs
&&
\nthsXvar{1}\A{a}{} = 
\atens{a}{bc}{} \A{b}{} \sX{c} 
+ \btens{a}{b'c} {*\linH{b'}} \sX{c} 
+ \ctens{a}{b'c} {*(\B{b'}{}} d\sX{c})
+ \tctens{a}{b'c} {*(*\B{b'}{} d\sX{c})} , 
\label{linsXAterms}\\
&&
\nthsXvar{1}\B{a'}{} = 
\dtens{a'}{b'c} \B{b'}{} \sX{c}
+ \tdtens{a'}{b'c} {*\B{b'}{}} \sX{c}
+\etens{a'}{bc} \linF{b} \sX{c}
+\tetens{a'}{bc} {*\linF{b}} \sX{c}
+\ftens{a'}{bc} {*(\A{b}{} d\sX{c})} , 
\label{linsXBterms}
\endEQs
with 
\EQ
\etens{a'}{[bc]} = \ftens{a'}{[bc]}=0 , 
\label{elin}
\endEQ
and by 
\EQs
\nthvXvar{1}\A{a}{} = &&
\gtens{a}{b'c'} {*(\B{b'}{} \vX{c'}{})}
+ \tgtens{a}{b'c'} {*(*\B{b'}{} \vX{c'}{})}
+\htens{a}{bc'} {*(\linF{b} \vX{c'}{})}
+\thtens{a}{bc'} {*(*\linF{b} \vX{c'}{})}
\nonumber\\&& 
+\itens{a}{bc'} \A{b}{} {*d*}\vX{c'}{}
+\titens{a}{bc'} (*d{*\A{b}{}}) \vX{c'}{} , 
\label{linvXAterms}\\
\nthvXvar{1}\B{a'}{} = &&
\jtens{a'}{bc'} \A{b}{} \vX{c'}{}
+ \tjtens{a'}{bc'} {*(\A{b}{} \vX{c'}{})}
+\ktens{a'}{b'c'} {*\linH{b'}} \vX{c'}{}
+\tktens{a'}{b'c'} {*(*\linH{b'} \vX{c'}{})}
+\ltens{a'}{b'c'} {*d*}(\B{b'}{} \vX{c'}{})
\nonumber\\&&
+\tltens{a'}{b'c'} {*d*}(*\B{b'}{} \vX{c'}{})
+\mtens{a'}{bc'} \B{b'}{} {*d*}\vX{c'}{}
+\tmtens{a'}{bc'} {*\B{b'}{}} {*d*}\vX{c'}{} , 
\label{linvXBterms}
\endEQs
where the coefficients are constants, 
which represent the components of bilinear maps from 
$\vsA \times\vsA$, $\vsA \times\vsB$, $\vsB \times\vsA$, $\vsB\times\vsB$
into $\vsA$ and $\vsB$. 
These coefficients are determined by 
solving the zeroth-order part of the Lie derivative commutator equation
from Theorem~2 (using the methods of \Ref{AMSpaper,Annalspaper}).
This yields the linear algebraic relations
\EQs
&&
\atens{}{a(bc)}{} =0 ,\quad
\ftens{}{a'(bc)} =0 , 
\\&&
\ltens{}{a'b'c'} = \tltens{}{a'b'c'} 
= \mtens{}{a'b'c'} = \tmtens{}{a'b'c'} =0 ,\quad
\gtens{}{ab'c'} = \tgtens{}{ab'c'} = 0 , 
\\&&
\ctens{}{ab'c} = \tctens{}{ab'c} = 0 ,\quad
\tjtens{}{a'bc'} = \tdtens{}{a'cb'} =0 ,\quad
\dtens{}{a'b'c} + \jtens{}{a'cb'} =0 .
\endEQs
Additional linear algebraic relations arise from 
the first-order part of the Lie derivative equation from Theorem~1
applied to the rigid symmetries 
\EQ
(\sXvar)_{\rm rigid} = \sXvar |_{d\sX{}=0} ,\quad
(\vXvar)_{\rm rigid} = \vXvar |_{d\vX{}{}=0} 
\endEQ
given by restricting the gauge symmetry variables 
so that $d\sX{a}=d\vX{a'}{}=0$.
This leads to (by the methods of \Ref{AMSpaper,Annalspaper})
\EQs
&&
\tetens{}{a'bc} + \btens{}{ba'c} =0 ,
\\&&
\Tm{a}{a'} \jtens{}{a'cb'} - \m{b'}{b} \atens{}{bac} =0 ,\quad
\atens{}{(ab)c}{} - \Tm{(b}{b'} \btens{}{a)b'c}{}=0 ,\quad
\jtens{}{(a'|c|b')}{} - \m{(a'}{a} \btens{}{|a|b')c}{} =0 , 
\label{jablin}\\
&&
\ktens{}{(a'b')e'}{} =0 ,\quad
\Tm{a}{a'} \ktens{}{a'b'c'}{} + \jtens{}{b'ac'}{} =0 ,
\label{kjlin}\\
&&
\htens{}{abc'} = \thtens{}{abc'} = 0 ,\quad
\tktens{}{a'b'c'} = 0 .
\endEQs
Then, we return to the first-order part of the Lie derivative equation 
with $\sX{a}$ and $\vX{a'}{}$ now taken to be arbitrary,
which determines the first-order deformation terms in the field equations
(by the methods of \Ref{JMPpaper1}). 
This yields
\EQs
* \nthEA{2}{a} = &&
{d*}( \frac{1}{2} \atens{a}{bc} \A{b}{} \A{c}{} 
+\btens{a}{b'c} {*\linH{b'}} \A{c}{} )
- \Tatens{cb}{a} \A{b}{} {*\linF{c}}
- \Tbtens{cb'}{a}{} {*\linH{b'}} \linF{c} 
\nonumber\\&&
+( 2\linF{b} {*\linH{c'}} -\A{b}{} d{*\linH{c'}} ) \etens{a'}{bc}
+\Tjtens{b'}{a}{c'} {*\linH{b'}} \B{c'}{}
- \m{c'}{c} \Tatens{cb}{a} \A{b}{} \B{c'}{} , 
\label{quadEA}\\
* \nthEB{2}{a'} = &&
{d*}( \jtens{a'}{bc'} \A{b}{} \B{c'}{} 
-\ktens{a'}{b'c'} {*\linH{b'}} \B{c'}{}
+\Tbtens{b}{a'}{c} {*\linF{b}} \A{c}{} )
+\frac{1}{2} \Tktens{b'c'}{a'} {*\linH{b'}} {*\linH{c'}}
\nonumber\\&&
-d( \A{b}{} {*\linF{c}} ) \etens{a'}{bc}
-\Tjtens{c'b}{a'}{} \A{b}{} {*\linH{c'}}
+\Tm{a}{a'} \atens{a}{bc} \A{b}{} \A{c}{} ,
\label{quadEB}
\endEQs
together with the linear algebraic relation 
\EQ
\Tm{(a|}{a'}\etens{}{a'b|c)} + \Tm{b}{b'}\etens{}{b'ac} =0 .
\label{emlin}
\endEQ
The corresponding Lagrangian is given by 
\EQs
\nthL{3} = &&
\frac{1}{2} \atens{}{abc} {*\linF{a}} \A{b}{} \A{c}{} 
+ \jtens{}{a'bc'} {*\linH{a'}} \A{b}{} \B{c'}{} 
+\btens{}{ab'c} {*\linF{a}} {*\linH{b'}} \A{c}{}
-\frac{1}{2} \ktens{}{a'b'c'} {*\linH{a'}} {*\linH{b'}} \B{c'}{} 
\nonumber\\&&
- \etens{}{a'bc} {*\linH{a'}} \linF{b} \A{c}{} 
+\frac{1}{2} \m{aa'}{} \atens{a}{bc} \B{a'}{} \A{b}{} \A{c}{} .
\label{cubicL}
\endEQs
These deformation terms are related to the deformation terms 
in the gauge symmetries 
(as follows from general results in \Ref{AMSpaper})
by being the Noether currents of the rigid symmetries 
associated to the first-order deformed gauge symmetries
\EQs
&&
\nthsXvar{1}\A{a}{} 
= \atens{a}{bc}{} \A{b}{} \sX{c} 
+ \btens{a}{b'c} {*\linH{b'}} \sX{c} , 
\label{linsXA}\\
&&
\nthsXvar{1}\B{a'}{} 
= -\jtens{a'}{cb'} \B{b'}{} \sX{c}
-\Tbtens{b}{a'}{c} {*\linF{b}} \sX{c}
+\etens{a'}{bc} \linF{b} \sX{c} , 
\label{linsXB}\\
&&
\nthvXvar{1}\A{a}{} 
= 0 , 
\label{linvXA}\\
&&
\nthvXvar{1}\B{a'}{} 
= \jtens{a'}{bc'} \A{b}{} \vX{c'}{}
+\ktens{a'}{b'c'} {*\linH{b'}} \vX{c'}{} . 
\label{linvXB}
\endEQs
In particular, we have
$*d( \nthEA{2}{a}\sX{b}\id{ab}{} )
= (\nthsXvar{1}\A{a}{})_{\rm rigid} \intprod \nthcoEA{1}{a}$
and 
$*d( \nthEB{2}{a'}\vX{b'}{}\id{a'b'}{} )
= (\nthsXvar{1}\B{a'}{})_{\rm rigid} \intprod \nthcoEB{1}{a'}$
where $d\sX{b}=d\vX{b'}{}=0$. 

We next note that, from \eqsref{linsXA}{linvXB}, 
\EQs
&& 
\nthcommXvar{0}{\sXsub{1}{}}{\sXsub{2}{}} = \nthsXvarsub{0}{3} ,\quad
\sXsub{3}{a} = \atens{a}{bc} \sXsub{1}{b} \sXsub{2}{c} , 
\label{lincommssX}\\
&&
\nthcommXvar{0}{\vXsub{1}{}{}}{\vXsub{2}{}{}} = 0 , 
\label{lincommvvX}\\
&&
\nthcommXvar{0}{\sXsub{1}{}}{\vXsub{2}{}{}} = \nthvXvarsub{0}{3} ,\quad
\vXsub{3}{a'}{} = \jtens{a'}{bc'} \sXsub{1}{b} \vXsub{2}{c'}{} .
\label{lincommsvX}
\endEQs

Now we consider 
the first-order part of the Lie derivative commutator equation from Theorem~2
and subtract the first-order part of the Lie derivative equation 
from Theorem~1 with the gauge symmetry variables 
given by the commutators \eqsref{lincommssX}{lincommsvX}. 
This combined equation leads to the result 
(by the methods of \Ref{AMSpaper,Annalspaper})
that the gauge symmetry commutators are closed to first-order 
when $\A{a}{},\B{a'}{}$ satisfy the linear field equations
$\nthEA{1}{a} =0$ and $\nthEB{1}{a'}=0$. 
Then if the gauge symmetry variables are taken to be rigid, 
$d\sXsub{1}{a} =d\sXsub{2}{a} =0$ 
and $d\vXsub{1}{a'}{}= d\vXsub{2}{a'}{} =0$,
we obtain an integrability condition
involving just the first-order deformation terms, 
\EQs 
&&
([ \nthsXvarsub{1}{1}, \nthsXvarsub{1}{2} ])_{\rm rigid} \A{a}{}
- (\nthsXvarsub{1}{3})_{\rm rigid} \A{a}{}
= d\nthsXsub{1}{a}{(\sXsub{1}{},\sXsub{2}{})} ,\quad
([ \nthsXvarsub{1}{1}, \nthsXvarsub{1}{2} ])_{\rm rigid} \B{a'}{}
- (\nthsXvarsub{1}{3})_{\rm rigid} \B{a'}{}
= d\nthvXsub{1}{a'}{(\sXsub{1}{},\sXsub{2}{})} , 
\\
&&
([ \nthvXvarsub{1}{1}, \nthvXvarsub{1}{2} ])_{\rm rigid} \A{a}{}
= d\nthsXsub{1}{a}{(\vXsub{1}{}{},\vXsub{2}{}{})} ,\quad
([ \nthvXvarsub{1}{1}, \nthvXvarsub{1}{2} ])_{\rm rigid} \B{a'}{}
= d\nthvXsub{1}{a'}{(\vXsub{1}{}{},\vXsub{2}{}{})} , 
\\
&&
([ \nthsXvarsub{1}{1}, \nthvXvarsub{1}{2} ])_{\rm rigid} \A{a}{}
- (\nthvXvarsub{1}{3})_{\rm rigid} \A{a}{}
= d\nthsXsub{1}{a}{(\sXsub{1}{},\vXsub{2}{}{})} ,\quad
([ \nthsXvarsub{1}{1}, \nthvXvarsub{1}{2} ])_{\rm rigid} \B{a'}{}
- (\nthvXvarsub{1}{3})_{\rm rigid} \B{a'}{}
= d\nthvXsub{1}{a'}{(\sXsub{1}{},\vXsub{2}{}{})} , 
\endEQs
which hold for some locally constructed
$\vsA$-valued functions $\nthsXsub{1}{a}{(\cdot,\cdot)}$
and $\vsB$-valued 1-forms $\nthvXsub{1}{a'}{(\cdot,\cdot)}$
depending linearly on 
$\sXsub{1}{a},\sXsub{2}{a},\vXsub{1}{a'}{},\vXsub{2}{a'}{}$. 
The solution of these six equations 
(using the methods of \Ref{Annalspaper}) 
yields the quadratic algebraic relations
\EQs
&&
\atens{}{adb} \atens{b}{ec} -2\atens{}{ab[c|} \atens{b}{d|e]} =0 , 
\label{aaquad}\\
&&
2\atens{}{ab[c} \btens{b}{|d'|e]} -\btens{}{ad'b} \atens{b}{ec}
+2\btens{}{ab'[c|} \btens{db'}{|e]} \m{d'd}{}
-2\btens{}{ab'[c|} \jtens{b'}{|e]d'} =0 , 
\label{abquad}\\
&&
\jtens{}{a'be'} \atens{b}{dc} -2\jtens{}{a'[d|b'} \jtens{b'}{|c]e'} =0 , 
\label{ajquad}
\endEQs
plus three others that are redundant 
as a consequence of \eqrefs{jablin}{kjlin}.

Another integrability condition arises 
for the first-order deformation terms
if we consider the second-order part of the Lie derivative equation 
from Theorem~1
under the previous conditions imposed on 
$\A{a}{},\B{a'}{},\sX{a},\vX{a'}{}$. 
Contracting this equation with $(\A{a}{},\B{a'}{})$,
we obtain
\EQs
&&
(\nthsXvar{1})_{\rm rigid} \A{a}{} \intprod \nthcoEA{2}{a} 
+ (\nthsXvar{1})_{\rm rigid} \B{a'}{} \intprod \nthcoEB{2}{a'} 
= d\nthstriv{3} , 
\\
&&
(\nthvXvar{1})_{\rm rigid} \A{a}{} \intprod \nthcoEA{2}{a} 
+ (\nthvXvar{1})_{\rm rigid} \B{a'}{} \intprod \nthcoEB{2}{a'} 
= d\nthvtriv{3} , 
\endEQs
holding for some locally constructed 3-forms
$\nthstriv{3},\nthvtriv{3}$
which depend linearly on 
$\sX{a},\vX{a'}{}$. 
The solution of these two equations 
(again using the methods of \Ref{Annalspaper}) 
yields the additional quadratic algebraic relations
\EQs
&& 
\ktens{}{[a'b'|c'} \ktens{c'}{|d']e'} =0 , 
\label{kkquad}\\
&& 
\btens{}{ab'c} \Tktens{d'e'}{b'} - 2\btens{}{a[d'|b} \btens{b}{|e']c} =0 , 
\label{bkquad}\\
&&
2\etens{}{a'(b|c} \atens{c}{|d)e} 
-\etens{}{c'bd} \jtens{c'}{ea'} 
-2\etens{}{a'c(d} \m{b)b'}{}\btens{cb'}{e} =0 , 
\label{aequad}
\endEQs
plus others that reduce to combinations of 
\eqsref{aaquad}{ajquad} through \eqrefs{jablin}{kjlin}. 
Moreover, the quadratic relation \eqref{ajquad}
itself is a consequence of \eqrefs{kkquad}{kjlin}.

It can be shown that the integrability relations \eqsref{aaquad}{ajquad}
are necessary and sufficient to allow 
solving for the second order deformation terms in the gauge symmetries
from the first-order part of the Lie derivative commutator equation
in Theorem~2.  
The additional integrability relations \eqsref{kkquad}{aequad} 
are necessary in then solving 
the second-order part of the Lie derivative equation from Theorem~1
for the second order deformation terms in the field equations. 
However, it is found that a solution exists if and only if
the following additional algebraic relation holds
on the coefficients of the field equation deformation terms, 
\EQ
\Tktens{d'e'}{c'} \etens{}{c'ab}
-2 \etens{}{[d'|ca} \btens{c}{|e']b}
-2 \etens{}{[d'|cb} \btens{c}{|e']a} =0 .
\label{ekquad}
\endEQ
This relation imposes in effect a further integrability relation
on allowed first-order deformations. 

\Proclaim{ Theorem~3 }{
Up to a change of field variables and gauge symmetry variables, 
all first-order geometrical deformations are given by 
\Eqsref{linsXA}{linvXB}, \Eqrefs{quadEA}{quadEB}, 
with the coefficients satisfying 
the linear relations \eqrefs{jablin}{kjlin}
and the quadratic relations \eqsref{aaquad}{ajquad}
and \eqsref{kkquad}{ekquad}.
There are no further algebraic obstructions to the existence of
second-order geometrical deformations.
}

These first-order deformations have the following classification:
the $\atens{}{abc}$ terms represent 
a massless \YM/ self-coupling of $\A{a}{}$, 
the $\ktens{}{a'b'c'}$ terms represent 
a massless \FT/ self-coupling of $\B{a'}{}$,
and the $\btens{}{ab'c}$ terms represent 
an extended \FT/ coupling between $\A{a}{}$ and $\B{a'}{}$,
while the $\jtens{}{a'bc'}$ terms represent 
a Higgs type coupling of $\B{a'}{}$ to $\A{a}{}$
which is nontrivial only when the mass tensor $\m{aa'}{}$ is nonzero. 
The $\etens{}{a'bc}$ terms, in contrast, represent
a different type of coupling between $\A{a}{}$ and $\B{a'}{}$
unrelated to \YM/ and \FT/ type couplings. 
It is similar in form to the coupling known for
1-form and 2-form fields in extended supergravity theory
\cite{ChaplineManton,Brandt2}. 
Moreover, 
the deformation corresponding to the $\etens{}{a'bc}$ terms
is characterized by possessing opposite parity
compared to the parity of the other deformation terms. 
In particular, consider the parity operator $\parity$
defined by $d\parity=\parity d$, $*\parity=-\parity *$. 
If we assign even parity to $\A{a}{}$ and odd parity to $\B{a'}{}$,
\EQ
\parity\A{a}{} = \A{a}{} ,\quad
\parity\B{a'}{} = -\B{a'}{} ,
\endEQ
which thus determines 
\EQ
\parity {*\linF{a}} = -\linF{a} ,\quad
\parity {*\linH{a'}} = \linH{a'} ,
\endEQ
and 
\EQ
\parity\sX{a} = \sX{a} ,\quad
\parity\vX{a'}{} = -\vX{a'}{} , 
\endEQ
then it follows that all the deformation terms 
except for the $\etens{}{a'bc}$ terms have even parity.

To proceed, we now consider the uniqueness of 
the higher-order deformation terms 
determined by the first-order terms in Theorem~3. 
Let $\Delta\nthEA{k+1}{a}$, $\Delta\nthEB{k+1}{a'}$, 
$\Delta\nthsXvar{k}\A{a}{}$, $\Delta\nthvXvar{k}\A{a}{}$, 
$\Delta\nthsXvar{k}\B{a'}{}$, $\Delta\nthvXvar{k}\B{a'}{}$
denote the difference of any two deformations that agree 
up to some given order $k\ge 1$. 
Then the $k+1$st order part of the Lie derivative equation from Theorem~1
yields 
\EQ
\nthsXvar{0}\Delta\nthEA{k+2}{a} 
= \nthvXvar{0}\Delta\nthEA{k+2}{a}=0 ,\quad
\nthsXvar{0}\Delta\nthEB{k+2}{a'} 
= \nthvXvar{0}\Delta\nthEB{k+2}{a'}=0 .
\label{diffkthEAB}
\endEQ
Similarly, the $k$th order part of 
the Lie derivative commutator equation from Theorem~2
yields the result that, 
after a change of field variables and gauge symmetry variables, 
\EQs
&&
\nthsXvarsub{0}{2}\Delta\nthsXvarsub{k+1}{1} \A{a}{}
= \nthvXvarsub{0}{2}\Delta\nthsXvarsub{k+1}{1} \A{a}{} =0 ,\quad
\nthsXvarsub{0}{2}\Delta\nthvXvarsub{k+1}{1} \A{a}{}
= \nthvXvarsub{0}{2}\Delta\nthvXvarsub{k+1}{1}\A{a}{} =0 , 
\label{diffkthXA}\\
&&
\nthsXvarsub{0}{2}\Delta\nthsXvarsub{k+1}{1} \B{a'}{}
= \nthvXvarsub{0}{2}\Delta\nthsXvarsub{k+1}{1} \B{a'}{} =0 ,\quad
\nthsXvarsub{0}{2}\Delta\nthvXvarsub{k+1}{1} \B{a'}{}
= \nthvXvarsub{0}{2}\Delta\nthvXvarsub{k+1}{1} \B{a'}{} =0 .
\label{diffkthXB}
\endEQs
Under the assumptions on the number of derivatives 
considered for possible deformation terms 
(see Remark in \secref{deteqs}), 
the solution of equations \eqsref{diffkthEAB}{diffkthXB} 
is immediately given by 
\EQs
&&
\Delta\nthEA{k+2}{a} =0 ,\quad
\Delta\nthEB{k+2}{a'} =0 ,\quad k\ge 1 , 
\\
&&
\Delta\nthsXvar{k+1}\A{a}{} = \Delta\nthvXvar{k+1}\A{a}{} =0 ,\quad
\Delta\nthsXvar{k+1}\B{a'}{} = \Delta\nthvXvar{k+1}\B{a'}{} =0 ,\quad k\ge 1 .
\endEQs
Hence, we have established the following uniqueness result. 

\Proclaim{ Theorem~4 }{
If two deformations agree at all orders $1\le l\le k$, 
$\Delta\nthEA{l+1}{a}=\Delta\nthEB{l+1}{a'}=0$, 
$\Delta\nthsXvar{l}\A{a}{}=\Delta\nthvXvar{l}\A{a}{}=0$, 
$\Delta\nthsXvar{l}\B{a'}{}=\Delta\nthvXvar{l}\B{a'}{}=0$,
then up to a change of field variables and gauge symmetry variables, 
the deformations also agree at order $l=k+1$. 
}

\subsection{ Deformations to all orders }

Hereafter we restrict attention to parity-invariant and opposite-parity
deformations separately
and proceed to write down a complete deformation to all orders
in each case. 
A full discussion of the combined parity non-invariant deformations
from Theorem~3 is given in \Ref{exotic}. 

For a deformation determined at first-order 
purely by the $\etens{}{a'bc}$ terms, 
note that the linear algebraic relations \eqrefs{elin}{emlin}
imply 
\EQ
\m{a'}{a} \etens{a'}{bc}=0
\endEQ
and hence the $\etens{}{a'bc}$ terms are incompatible
with a nonzero mass tensor.
However, in the massless case, 
the $\etens{}{a'bc}$ terms produce a nontrivial deformation,
which we now write down to all orders. 

The algebraic structure on $\vsA,\vsB$ associated to $\etens{a'}{bc}$
consists of a symmetric product from $\vsA \times\vsA$ into $\vsB$. 
Then the gauge symmetries are given by 
\EQs
&&
\sXvar\A{a}{} =d\sX{a} ,\quad
\sXvar\B{a'}{} =\etens{a'}{bc} \linF{b} \sX{c} , 
\label{sXvarnew}\\
&&
\vXvar\A{a}{} =0 ,\quad
\vXvar\B{a'}{} =d\vX{a'}{} , 
\label{vXvarnew}
\endEQs
while the Lagrangian is constructed by 
\EQ
L 
= \frac{1}{2} \linF{a} {*\linF{b}} \id{ab}{}
- \frac{1}{2} \linH{a'} {*\linH{b'}} \id{a'b'}{}
-*\linH{a'} \linF{b} \A{c}{} \etens{}{a'bc}
+ \frac{1}{2} \linF{b} \A{c}{} {*( \linF{d} \A{e}{} )} 
\etens{a'}{bc} \etens{}{a'de} . 
\label{Lnew}
\endEQ
It is straightforward to see that this Lagrangian is gauge invariant, 
$\sXvar L = \vXvar L =0$, 
and that the gauge symmetries commute, 
$[\sXvarsub{1},\sXvarsub{2}] 
= [\vXvarsub{1},\vXvarsub{2}] 
= [\sXvarsub{1},\vXvarsub{2}] 
=0$. 
From the Lagrangian, the field equations are given by 
\EQs
\EA{a}{} = &&
{d*}\linF{a} 
+ ( 2\linF{b} {*\linH{c'}} - \A{b}{} d{*\linH{c'}} )\Tetens{c'b}{a}
\nonumber\\&& 
-(2 \linF{b} {*(\linF{c}\A{d}{})} + \A{b}{} {d*}(\linF{c}\A{d}{}) )
\etens{a'}{cd} \Tetens{a'b}{a} =0 , 
\label{AEnew}\\
\EB{a'}{} = &&
d( *\linH{a'} -{*(\linF{b}\A{c}{})} )\etens{a'}{bc} =0 . 
\label{BEnew}
\endEQs

\Proclaim{ Theorem~5 }{
The massless nonlinear theory \eqsref{sXvarnew}{BEnew}
is the unique nonlinear geometrical deformation of
the abelian linear theory \eqsref{quadL}{linEB}
determined by the first-order deformation terms $\etens{}{a'bc}$. 
}

Next we consider a general deformation determined at first-order 
by all terms except $\etens{}{a'bc}$. 
This deformation is more general than 
the massless/massive nonlinear theories 
constructed in \secrefs{masslessth}{massiveth},
since it includes a mixing of massless and massive fields
$\A{a}{},\B{a'}{}$, 
controlled by the eigenvalues of the mass tensor $\m{aa'}{}$. 

Let $\atens{a}{bc},\btens{a}{b'c},\jtens{a'}{bc'},\Tktens{a'b'}{c'}$
be the components of respective bilinear maps from 
$\vsA\times\vsA$ into $\vsA$, 
$\vsB\times\vsA$ into $\vsA$,
$\vsA\times\vsB$ into $\vsB$,
$\vsB\times\vsB$ into $\vsB$, 
fixed to satisfy the linear and quadratic relations 
\eqrefs{jablin}{kjlin}, \eqsref{aaquad}{ajquad}, \eqrefs{kkquad}{bkquad}. 
Thus, it follows that 
$\atens{a}{bc}$ and $\Tktens{a'b'}{c'}$ 
define the commutator structure constants \eqrefs{aaquad}{kkquad} 
of respective Lie algebras on $\vsA,\vsB$,
while $\jtens{a'}{bc'}$ and $\btens{a}{b'c}$ 
define linear maps that are representations \eqrefs{ajquad}{bkquad}
of these Lie algebras 
on the vector spaces of $\vsB,\vsA$, respectively. 
Further discussion of the additional algebraic structure imposed by 
the relations \eqref{jablin}, \eqref{kjlin}, \eqref{abquad} 
is given in the next subsection. 

To write down the deformation to all orders, 
we first define a \YM/ field strength 2-form 
and a related antisymmetric tensor field strength 3-form by 
\EQ
\F{a}{A} =d\A{a}{} +\frac{1}{2} \atens{a}{bc} \A{b}{} \A{c}{} ,\quad
\H{a'}{A} = d\B{a'}{} +\jtens{a'}{bc'}\A{b}{} \B{c'}{} . 
\label{nonlinFH}
\endEQ
Geometrically, 
$\F{a}{A}$ is the curvature of the connection 1-form 
$\atens{a}{bc} \A{b}{}$,
and $\H{a'}{A}$ is the covariant curl of $\B{a'}{}$
in terms of the associated connection 
$\jtens{a'}{bc'}\A{b}{}$. 
Consequently, using the covariant exterior derivative operators given by 
\EQ
\DA{A} =d + \atens{a}{bc} \A{b}{} ,\quad
\DB{A} =d + \jtens{a'}{bc'}\A{b}{} ,
\endEQ
we have 
\EQs
& \DB{A} \B{a'}{} = \H{a'}{A} ,&
\\
& [\DA{A},\DA{A}] = \atens{a}{bc} \F{b}{A} ,\quad
[\DB{A},\DB{A}] = \jtens{a'}{bc'} \F{b}{A} ,&
\endEQs
due to the algebraic structure \eqrefs{aaquad}{ajquad}. 
Next we define a nonlinear field strength 
2-form $\P{a}{}$ and 3-form $\Q{a}{}$ by the equations
\EQs
&&
\P{a}{} -*\Q{b'}{} \A{c}{} \btens{a}{b'c} 
= \F{a}{A} , 
\label{nonlinP}\\
&&
\Q{a'}{} -*\P{b}{} \A{c}{} \Tbtens{b}{a'}{c} 
-*\Q{b'}{} \B{c'}{} \ktens{a'}{b'c'}
= \H{a'}{A} .
\label{nonlinQ}
\endEQs
These field strengths are nonpolynomial expressions 
in terms of $\A{a}{},\B{a'}{}$, as given by 
\EQ
(\P{a}{},\Q{a'}{}) = \invY{}{A,B}(\F{a}{A},\H{a'}{B})
\endEQ
where $\invY{}{A,B}$ is the inverse of the linear map 
\EQ
\Y{}{A,B} = 
\matr{ \idmap & \A{b}{} \btens{a}{c'b}* \cr
 -\A{b}{} \Tbtens{c}{a'}{b}* & \idmap -\B{b'}{} \ktens{a'}{c'b'}* }
\endEQ
defined to act on the vector space of pairs of 
$\vsA$-valued 2-forms, $\vsB$-valued 3-forms. 

Now we write down the deformation, using the previous structure. 
The gauge symmetries on $\A{a}{},\B{a'}{}$ are given by 
the field variations 
\EQs
&&
\sXvar\A{a}{} 
= \DA{A} \sX{a} + \btens{a}{b'c} {*\Q{b'}{}} \sX{c} , 
\label{nonlinsXA}\\
&&
\sXvar\B{a'}{} 
= -\jtens{a'}{cb'} \B{b'}{} \sX{c} -\Tbtens{b}{a'}{c} {*\P{b}{}} \sX{c} , 
\label{nonlinsXB}\\
&&
\vXvar\A{a}{} 
= 0 , 
\label{nonlinvXA}\\
&&
\vXvar\B{a'}{} 
= \DB{A} \vX{a'}{} + \ktens{a'}{b'c'} {*\Q{b'}{}} \vX{c'}{} , 
\label{nonlinvXB}
\endEQs
where $\sX{a}$ is an arbitrary $\vsA$-valued function, 
and $\vX{a'}{}$ is an arbitrary $\vsB$-valued 1-form. 
The Lagrangian 4-form for $\A{a}{},\B{a'}{}$ is constructed by 
\EQ
L = 
\frac{1}{2} {*\P{a}{}} \F{b}{A} \id{ab}{} 
+ \frac{1}{2} {*\Q{a'}{}} \H{b'}{A} \id{a'b'}{}
+\F{a}{A} \B{b'}{} \m{ab'}{} . 
\label{nonlinL}
\endEQ
Gauge invariance of this Lagrangian is established as follows. 

The variation of $L$ under the gauge symmetry $\sXvar$ yields
\EQs
\sXvar L = &&
\m{ab'}{} (\DA{A} \sXvar\A{a}{} \B{b'}{} +\F{a}{A} \sXvar\B{b'}{} )
+ {*\P{e}{}}( \DA{A} \sXvar\A{a}{} 
+ \btens{a}{b'c} {*\Q{b'}{}} \sXvar\A{c}{} )\id{ae}{}
\nonumber\\&&
+ {*\Q{e'}{}}( \DB{A} \sXvar\B{a'}{} 
+ \jtens{a'}{bc'} \sXvar\A{b}{} \B{c'}{} 
+ \frac{1}{2} \ktens{a'}{b'c'} {*\Q{b'}{}} \sXvar\B{c'}{} )\id{a'e'}{} . 
\endEQs
To proceed, 
collecting all terms $d\sX{c}$ we obtain 
$-( *\P{b}{} \btens{}{ba'c} +\B{b'}{} \m{bb'}{} \btens{b}{a'c} )
*\Q{a'}{} d\sX{c}$. 
Next we integrate by parts and use the field strength equations
\eqref{nonlinFH}, \eqref{nonlinP}, \eqref{nonlinQ} 
to eliminate $d\A{a}{},d\B{a'}{}$ algebraically in terms of 
$\P{a}{},\Q{a'}{},\A{a}{},\B{a'}{}$. 
This yields terms of the type
$\P{a}{}{*\P{b}{}}\sX{c}$, $\Q{a'}{}*\Q{b'}{}\sX{c}$, 
$*\Q{a'}{}*\Q{b'}{}*\P{c}{}\sX{d}$, 
$*\P{a}{}*\Q{b'}{}\A{c}{}\sX{d}$, 
$*\Q{a'}{}*\Q{b'}{}\B{c'}{}\sX{d}$. 
Then we find that the coefficients of these terms vanish, respectively, 
due to the algebraic relations
\eqref{jablin}, \eqref{bkquad}, \eqref{abquad}, \eqref{ajquad}. 
Hence, it follows that 
\EQ
\sXvar L =
d( (\btens{}{ba'c} \P{b}{} +\m{bb'}{} \btens{b}{a'c} \B{b'}{}) 
{*\Q{a'}{}} \sX{c} ) . 
\label{sXvarL}
\endEQ

Similarly, under the gauge symmetry $\vXvar$, 
the variation of $L$ is given by 
\EQs
\vXvar L = &&
\m{ab'}{} \F{a}{A} \vXvar\B{b'}{} 
+ {*\Q{e'}{}}( \DB{A} \vXvar\B{a'}{} 
+ \frac{1}{2}\ktens{a'}{b'c'} {*\Q{b'}{}} \vXvar\B{c'}{} )\id{a'e'}{} . 
\endEQs
Proceeding as before, we find that all $d\vX{c'}{}$ terms yield
$( -\frac{1}{2} {*\Q{a'}{}} {*\Q{b'}{}} \ktens{}{a'b'c'} 
+\F{a}{A} \m{ac'}{} )d\vX{c'}{}$. 
An integration by parts and use of the field strength equations
\eqref{nonlinFH}, \eqrefs{nonlinP}{nonlinQ}
leaves terms of the type
$*\Q{a'}{}\P{b}{}\vX{c'}{}$, 
$*\Q{a'}{} {*\Q{b'}{}} {*\Q{c'}{}} \vX{d'}{}$, 
$*\Q{a'}{} {*\Q{b'}{}} \A{c}{}\vX{d'}{}$, 
$\P{a}{}\A{b}{}\vX{c'}{}$, 
$\A{a}{}\A{b}{} {*\Q{c'}{}} \vX{d'}{}$. 
Then we find that the coefficients of these terms vanish, respectively, 
due to the algebraic relations
\eqref{kjlin}, \eqref{kkquad}, \eqref{ajquad}, \eqref{jablin}. 
Hence, it follows that 
\EQ
\vXvar L =
d( (\frac{1}{2}\ktens{}{a'b'c'} {*\Q{a'}{}} {*\Q{b'}{}}
+\m{ac'}{}\F{a}{A}) \vX{c'}{} . 
\label{vXvarL}
\endEQ

\Proclaim{ Proposition }{
The Lagrangian \eqref{nonlinL} is invariant 
to within an exact 4-form \eqrefs{sXvarL}{vXvarL} 
under the gauge symmetries \eqsref{nonlinsXA}{nonlinvXB}. }

The field equations for $\A{a}{},\B{a'}{}$ are given by 
\EQs
&&
\EA{a}{} 
= \DA{A} {*\P{a}{}}
+(\btens{a}{c'b}\Tm{c}{c'} \A{b}{} -\Tbtens{cb'}{a}{} {*\Q{b'}{}}) {*\P{c}{}}
+ \Q{b'}{} \m{b'}{a} 
=0 , 
\label{nonlinEA}\\
&&
\EB{a'}{} 
= \DB{A} {*\Q{a'}{}}
+\frac{1}{2} \Tktens{b'c'}{a'} {*\Q{b'}{}} {*\Q{c'}{}}
-\Tbtens{c}{a'}{b} \m{c'}{c} \A{b}{} {*\Q{c'}{}}
+ \P{b}{} \Tm{b}{a'} 
=0 .
\label{nonlinEB}
\endEQs
On solutions of these field equations, 
the gauge symmetries on $\A{a}{},\B{a'}{}$ have the commutator structure
\EQ
[\sXvarsub{1},\sXvarsub{2}] = \sXvarsub{3} ,\quad
[\vXvarsub{1},\vXvarsub{2}] = 0 ,\quad
[\sXvarsub{1},\vXvarsub{2}] = \vXvarsub{3} 
\endEQ
where
\EQ
\sXsub{3}{a} = \atens{a}{bc} \sXsub{1}{b} \sXsub{2}{c} ,\quad
\vXsub{3}{a'}{} = \jtens{a'}{bc'} \sXsub{1}{b} \vXsub{2}{c'}{} .
\endEQ
Off solutions, the commutators structure remains closed
to within trivial symmetries proportional to the field equations
\EQs
\delta_E\A{a}{} = &&
2\btens{ab'}{[c|} \btens{}{db'|e]} \sXsub{1}{c} \sXsub{2}{e} {*\EA{d}{}}
- \btens{ab'}{c} \ktens{}{b'd'e'} \sXsub{1}{c} 
{*( \vXsub{2}{e'}{} \EB{d'}{} )} ,
\\
\delta_E\B{a'}{} = &&
2\btens{a'b}{[c|} \btens{}{bd'|e]} \sXsub{1}{c} \sXsub{2}{e} {*\EB{d'}{}}
+ \ktens{a'b'}{e'} \btens{}{db'c} \sXsub{1}{c} \vXsub{2}{e'}{} {*\EA{d}{}}
\nonumber\\&&
- \ktens{a'b'}{c'} \ktens{}{b'd'e'} \vXsub{1}{c'}{} 
{*( \vXsub{2}{e'}{} \EB{d'}{} )}
+ \ktens{a'b'}{e'} \ktens{}{b'd'c'} \vXsub{2}{e'}{} 
{*( \vXsub{1}{c'}{} \EB{d'}{} )} .
\endEQs

\Proclaim{ Theorem~6 }{
The nonlinear theory \eqsref{nonlinsXA}{nonlinEB}
is the unique nonlinear geometrical deformation of
the abelian linear theory \eqsref{quadL}{linEB}
determined by the first-order deformation terms 
$\atens{a}{bc},\btens{a}{b'c},\jtens{a'}{bc'},\ktens{a'}{b'c'}$.
}

We remark that the pure massless/massive $\SU{2}$ case of this theory, 
given by 
\EQ
\atens{a}{bc} =\cross{bc}{a} ,\quad
\btens{a}{b'c} =\lambda \cross{b'c}{a} ,\quad
\jtens{a'}{bc'} =\cross{bc'}{a'} ,\quad
\ktens{a'}{b'c'} =\lambda \cross{b'c'}{a'} 
\endEQ
(where $a,a',\ldots=1,2,3$)
with $\lambda =1/m$ and $m=\const\neq 0$ in the massive case, 
and $\lambda=\const\neq 0$ in the massless case, 
yields the $\SU{2}$ theories from \secrefs{masslessth}{massiveth}.

\subsection{ Algebraic structure in the nonlinear theory } 

Finally, we discuss the full algebraic structure on $\vsA,\vsB$
underlying the nonlinear theory \eqsref{nonlinsXA}{nonlinEB}
given by the general parity-invariant deformation. 

We start from the vector space decompositions
$\vsA = \vsA_0 \oplus \vsA_{\rm m}$
and $\vsB=\vsB_0 \oplus \vsB_{\rm m}$
into massless and massive subspaces defined by the mass tensor $\m{aa'}{}$. 
Let $\projA,\mprojA,\projB,\mprojB$ 
be the respective projection operators onto these subspaces in $\vsA,\vsB$.
Thus we have $\mA(u_{0})=\mB(v'_{0})=0$
and $u'_{\rm m}=\mA(u_{\rm m})\neq 0$, 
$v_{\rm m}=\mB(v'_{\rm m})\neq 0$,
with subscripts denoting subspace projections 
for all $u,v$ in $\vsA$ and $u',v'$ in $\vsB$. 
Let $(u,v)_\vsA=\u{a}{} \v{b}{} \id{ab}{}$
and $(u',v')_\vsB=\uB{a}{} \vB{b}{} \id{a'b'}{}$
denote the vector space inner products on $\vsA,\vsB$.
We denote the Lie algebra multiplication 
$\atens{a}{bc} \u{b}{} \v{c}{}$ and $\Tktens{b'c'}{a'} \uB{b}{} \vB{c}{}$ 
on $\vsA$ and $\vsB$ by the brackets $[u,v]_\vsA$ and $[u',v']_\vsB$. 
In addition we denote the Lie algebra representations 
$\jtens{a'}{bc'}\u{b}{}$ and $\btens{a}{b'c}\uB{b}{}$
by the linear maps $\reprB(u)$ and $\reprA(u')$. 
Similarly, $\adA(u)$ and $\adB(u')$
denote the adjoint representations given by 
$\atens{a}{bc} \u{b}{}$ and $\Tktens{b'c'}{a'} \uB{b}{}$
on $\vsA$ and $\vsB$. 

We begin by noting \Eqrefs{kkquad}{bkquad} show that
$\reprB$ is a derivation of the Lie algebra $\vsB$
\EQ
\reprB(w)[u',v']_\vsB = [\reprB(w)u',v']_\vsB + [u',\reprB(w)v']_\vsB 
\endEQ
for all $w$ in $\vsA$, $u'v'$ in $\vsB$. 

We now consider the additional algebraic structure imposed by 
the algebraic relations \eqrefs{jablin}{kjlin}. 
To proceed, first note that, by \Eqref{kjlin}, 
\EQ
\reprB(u_{0}) =0 ,\quad
\reprB(u_{\rm m}) =\adB(\mA(u_{\rm m})) , 
\endEQ
which completely determines $\reprB$ 
in terms of the adjoint representation of $\vsB$. 
Next, by \Eqref{jablin}, it follows that
\EQ
\mA([u,v]_\vsA) =[\mA(u),\mA(v)]_\vsB . 
\label{BAhomomorphism}
\endEQ
Hence, 
$\vsB_{\rm m}$ is a Lie subalgebra of $\vsB$,
and $\vsA_{0}$ is an invariant Lie subalgebra of $\vsA$, 
namely 
$[\vsB_{\rm m},\vsB_{\rm m}] \subseteq \vsB_{\rm m}$, 
$[\vsA,\vsA_{0}] \subseteq \vsA_{0}$. 
Furthermore, it also follows from \Eqref{jablin} that
\EQ
(u_{0},[v_{0},w]_\vsA)_\vsA = - (v_{0},[u_{0},w]_\vsA)_\vsA 
\endEQ
and thus the inner product on $\vsA$ is an invariant metric
with respect to the massless subspace $\vsA_0$.
Consequently, since $\vsA_0$ is a Lie subalgebra, 
it must be a direct sum of an abelian Lie algebra $\vsA_0^c$
and a semisimple Lie algebra $\vsA_0^s$. 
However, the inner product is not required to be invariant 
with respect to the whole Lie algebra $\vsA$, since
\EQ
(u,[v,w]_\vsA)_\vsA + (v,[u,w]_\vsA)_\vsA 
= (u,\reprA(\mA(v))w)_\vsA + (v,\reprA(\mA(u))w)_\vsA 
\endEQ
which need not vanish for $u,v$ in $\vsA_{\rm m}$.
Thus, surprisingly, $\vsA$ need not be semisimple
unless its massive subspace $\vsA_{\rm m}$ is empty. 
Moreover, 
the inner product on the Lie algebra $\vsB$ 
likewise is not required to be invariant
except on the massless subspace $\vsB_0$
\EQ
(u'_{0},[v'_{0},w']_\vsB)_\vsB = - (v'_{0},[u'_{0},w']_\vsB)_\vsB , 
\label{invmetricB}
\endEQ
since, by \Eqref{kjlin}, 
\EQ
(u',[v',\mA(w)]_\vsB)_\vsB + (v',[u',\mA(w)]_\vsB)_\vsB 
= (u',\mA(\reprA(v')w))_\vsB + (v',\mA(\reprA(u')w))_\vsB 
\endEQ
which need not vanish for $u',v'$ in $\vsB_{\rm m}$.
Hence, as $\vsB_{\rm m}$ is a Lie subalgebra,
it need not be semisimple
and therefore, again, 
the whole Lie algebra $\vsB$ is not required to be semisimple
unless its massive subspace $\vsB_{\rm m}$ is empty. 

Finally, we consider the remaining algebraic relation \eqref{abquad}. 
This imposes further structure on the Lie algebras $\vsA,\vsB$,
and on the representation $\reprA$ as follows. 
We first examine, separately, 
the pure massless case $\m{aa'}{}=0$
and pure massive case $\m{aa'}{}=m\id{aa'}{}$, $m\neq 0$.

In the massless case, 
\EQ
\vsA =\vsA_0 ,\quad
\vsB =\vsB_0 , 
\endEQ
and so $\vsA$ and $\vsB$ are each a direct sum of 
abelian Lie algebras $\vsA^c,\vsB^c$
and semisimple Lie algebras $\vsA^s,\vsB^s$, respectively. 
Now, \Eqref{abquad} reduces to 
\EQ
\reprA(w')[u,v]_\vsA = [\reprA(w')u,v] + [u,\reprA(w')v] , 
\label{abeqmassless}
\endEQ
which states that the linear map $\reprA$ is 
a derivation of the Lie algebra $\vsB$. 
Since the quotient of $\vsB$ by its center $\vsB^c$ is semisimple, 
$\reprA$ must take the form
\EQ
\reprA(w') =\adA(\BintoA(w'))
\endEQ
for some linear map $\BintoA$ from $\vsB$ into $\vsA$. 
It then follows that
\EQ
[\BintoA(u'),\BintoA(v')]_\vsA = \BintoA([u',v']_\vsB) . 
\endEQ
Hence, 
the kernel of $\BintoA$ is an invariant Lie subalgebra of $\vsB$.
As a consequence of the decomposition $\vsB=\vsB^c \oplus \vsB^s$, 
any such subalgebra must belong to $\vsB^c$. 
Furthermore, 
the image of $\BintoA$ is a Lie subalgebra of $\vsA$
isomorphic to $\vsB^s$.
Hence, $\BintoA(\vsB^s) \subseteq \vsA^s$
yields a Lie algebra homomorphism. 
This now fully describes the structure imposed on $\vsA$ and $\vsB$
by the algebraic relations 
\eqrefs{jablin}{kjlin}, \eqsref{aaquad}{ajquad}, \eqrefs{kkquad}{bkquad}
in the massless case. 

In the massive case, note 
\EQ
\vsA =\vsA_{\rm m} ,\quad
\vsB =\vsB_{\rm m}
\endEQ
are isomorphic as vector spaces under the map 
$\mB(\vsB_{\rm m}) = m \idmap(\vsB_{\rm m}) = m \vsA_{\rm m}$. 
Since this map is a Lie algebra homomorphism by \Eqref{BAhomomorphism}, 
then $\vsA$ and $\vsB$ are isomorphic Lie algebras. 
Consequently, \Eqref{abquad} becomes
\EQs
&&
[\reprAm(w)u,v]_\vsA + [u,\reprAm(w)v]_\vsA -\reprAm(w)[u,v]_\vsA 
\nonumber\\&&
= \reprAm( \reprAm(w)u )v - \reprAm( \reprAm(w)v )u
-\reprAm( [w,u]_\vsA )v + \reprAm( [w,v]_\vsA )u
\label{abeqmassive}
\endEQs
where $\reprAm(w) = \reprA(\mA(w))$. 
Taking into account \Eqref{jablin},
the relation \eqref{abeqmassive} states that
the linear map on $\vsA$ defined by 
\EQ
\treprAm(u)v = \reprAm(v)u -\adA(v)u
\endEQ
must be a skew-adjoint representation of the Lie algebra $\vsA$,
\EQ
(u,\treprAm(w)v)_\vsA = - (v,\treprAm(w)u)_\vsA ,\quad
[\treprAm(u),\treprAm(v)] = \treprAm([u,v]_\vsA) . 
\endEQ
This is satisfied if $\reprAm=\adA$,
in which case $\treprAm=0$ is a trivial representation, 
or if $\reprAm=0$, 
in which case $\treprAm=-\adTA$ is the coadjoint representation. 
In either case, 
there is no further algebraic structure imposed by \Eqref{abeqmassive}.
Note then, surprisingly, $\vsA$ is thus not required to be semisimple
in massive case with $\reprAm=\adA$. 

To conclude the discussion, 
we return to the general situation when $\vsA$ and $\vsB$ 
contain both massless and massive nonempty subspaces. 
In this case, from \Eqref{abquad}, it follows that
\EQs
&&
[\reprA(w')u,v]_\vsA + [u,\reprA(w')v]_\vsA -\reprA(w)[u,v]_\vsA 
\nonumber\\&&
= \reprA( \mA(\reprA(w')u) )v - \reprA( \mA(\reprA(w')v) )u
-\reprA( [w',\mA(u)]_\vsB )v + \reprA( [w',\mA(v)]_\vsB )u . 
\label{abeq}
\endEQs
We now show that this equation is satisfied by 
\EQ
\reprA(w') = \adA(\BintoA(w'))
\label{reprAsol}
\endEQ
for some linear map $\BintoA$ from $\vsB$ into $\vsA$
if 
\EQ
\BintoA(\mA(\vsA_{\rm m})) = \vsA_{\rm m} ,\quad
\mA(\BintoA(\vsB_{\rm m})) = \vsB_{\rm m} ,\quad
\BintoA(\vsB_{0}) \subseteq \vsA_{0} 
\label{propBintoA}
\endEQ
and if
\EQ
[\vsA_{\rm m},\vsA_{\rm m}] \subseteq \vsA_{\rm m} ,\quad
[\vsA_{\rm m},\vsA_{0}] \subseteq \vsA_{0}^c ,\quad 
[\adAzero(\vsA_{\rm m}),\adAzero(\vsA_{\rm m})] =0 
\label{propLiealgA}
\endEQ
where $\vsA_{0}^c$ is the center of the Lie algebra $\vsA$. 
To begin the proof, 
first note that if \Eqref{reprAsol} holds
then the left-side of \Eqref{abeq} directly vanishes for any $\BintoA$
since $\adA$ is a derivation of $\vsA$.
Next, from \Eqref{bkquad}, 
the last two terms on the right-side of \Eqref{abeq} become
\EQs
&&
-[\adA(w_{\rm m}),\adA(u_{\rm m})] v
+ [\adA(w_{\rm m}),\adA(v_{\rm m})] u
\nonumber\\&&
= -\adA([w_{\rm m},u_{\rm m}]_\vsA) v
+ \adA([w_{\rm m},v_{\rm m}]_\vsA) u
= -[[w_{\rm m},u_{\rm m}]_\vsA,v]_\vsA 
+ [[w_{\rm m},v_{\rm m}]_\vsA,v]_\vsA 
\endEQs
where $w_{\rm m}=\mB(w')$. 
Then, by \Eqref{propLiealgA}, 
the first two terms on the right-side of \Eqref{abeq} reduce to 
\EQ
\adA( \adA(w_{\rm m})u_{\rm m} )v 
- \adA( \adA(w_{\rm m})v_{\rm m} )u
= [[w_{\rm m},u_{\rm m}]_\vsA,v]_\vsA 
- [[w_{\rm m},v_{\rm m}]_\vsA,u]_\vsA  . 
\endEQ
Hence, the right-side of \Eqref{abeq} vanishes,
which completes the proof. 

Consequently, 
note that \Eqsref{reprAsol}{propLiealgA}
determine 
\EQ
\reprA(w'_{\rm m}) =\adAm(\minvAB(w'_{\rm m})) ,\quad
\reprA(w'_{0}) =\adAzero(h_0(w'_{0})) 
\endEQ
in terms of some linear map $h_0=\BintoA \circ \projB$
from $\vsB_0$ into $\vsA_0$,
and using the inverse $\minvAB$ from $\vsA_{\rm m}$ into $\vsB_{\rm m}$
of the linear map $\mB \circ \mprojB$. 
It then follows that 
\EQ
\mB([u',v']_\vsB) = [\mB(u'),\mB(v')]_\vsA . 
\label{BAmhomomorphism}
\endEQ
Hence, $\vsB_0$ is an invariant Lie subalgebra of $\vsB$,
and $\vsA_{\rm m}$ is a Lie subalgebra of $\vsA$.
Consequently, 
since by \Eqref{invmetricB} the inner product on $\vsB_0$
is an invariant metric with respect to $\vsB$,
the Lie subalgebra $\vsB_0$ is a direct sum of 
an abelian Lie algebra $\vsB_0^c$ and a semisimple Lie algebra $\vsB_0^s$. 
Furthermore, from \Eqref{BAmhomomorphism},
it follows that $\vsA_{\rm m}$ and $\vsB_{\rm m}$ 
are isomorphic Lie algebras, 
but note that they are not required to be semisimple. 
This now gives a complete description of the algebraic structure
imposed by the relations 
\eqrefs{jablin}{kjlin}, \eqsref{aaquad}{ajquad}, \eqrefs{kkquad}{bkquad}
in the case given by \Eqsref{reprAsol}{propLiealgA}. 

Thus, the previous algebraic analysis leads to 
an interesting generalization of the massless/massive nonlinear theories
in \secrefs{masslessth}{massiveth} 
given by the nonlinear theory \eqsref{nonlinsXA}{nonlinEB}
with the following algebraic structure:

(i) The massless and massive subspaces 
$\vsA_0,\vsB_0,\vsA_{\rm m},\vsB_{\rm m}$ 
are Lie subalgebras of $\vsA,\vsB$
with $\vsB_{\rm m}$ and $\vsA_{\rm m}$ being isomorphic 
under the linear maps $\mA,\mB$ given by the mass tensor. 

(ii) $\vsA_0$ and $\vsB_0$ are semisimple Lie algebras 
and ideals in $\vsA,\vsB$, 
such that $\vsA_0$ and $\vsA_{\rm m}$ commute; 
however, the Lie algebras $\vsA_{\rm m} \simeq \vsB_{\rm m}$ 
are not restricted to be semisimple (they may be nilpotent or solvable)
and $\vsB_{\rm m}$ is not restricted to commute with $\vsB_0$. 

(iii) The representation $\reprB$ is the adjoint representation of
$\vsB_{\rm m}=\mA(\vsA_{\rm m})$ on $\vsB$, 
while the representation $\reprA$ is the sum of 
the adjoint representations of $\vsA_{\rm m}=\minvAB(\vsB_{\rm m})$ 
and of $\vsA_{0}=h_0(\vsB_{0})$ on $\vsA$,
for any linear map $h_0$. 

In physical terms, 
the resulting nonlinear theory \eqsref{nonlinsXA}{nonlinEB}
is a novel generalization of 
\YM/ gauge theory for vector potentials $\A{a}{}$
coupled to \FT/ gauge theory for antisymmetric tensor potentials $\B{a'}{}$,
involving a \CS/ type mass term. 
It describes a set of nonlinearly interacting massive spin-one fields
and massless spin-one and spin-zero fields, 
with a mutual interaction between the massive and massless fields.

\section{ Concluding remarks }
\label{conclude}

This paper has developed in detail 
the geometrical, field theoretic, and algebraic aspects of
an interesting nonlinear generalization of 
massless/massive \YM//\FT/ gauge theory in four dimensions. 
The generalization involves an extended \FT/ coupling between
the \YM/ 1-form gauge fields and \FT/ 2-form gauge fields,
in addition to a Higgs type coupling tied to a \CS/ mass term,
and accompanied by a novel form of generalized \YM/ and \FT/ 
gauge symmetries and field equations 
in both the massless and massive cases. 
In particular, the geometrical structure of 
the resulting nonlinear gauge theory 
mixes and unifies well-known features of \YM/ theory and \FT/ theory
in terms of Lie algebra valued curvatures and connections associated to
the gauge fields and nonlinear field strengths. 

This generalization was found by a general determination of
the geometrical nonlinear deformations of 
linear abelian gauge theory for 1-form fields and 2-form fields
with an abelian \CS/ mass term. 
The deformation framework used here is a geometrical version of
the field theoretic approach developed in 
\Ref{AMSpaper,JMPpaper1,Annalspaper}.
It exposes clearly the existence of two integrability conditions
on the first-order parts of possible deformations
and leads to a simple uniqueness argument 
for the higher-order parts of allowed deformations. 

Another approach to deformations (see \Ref{Henneaux1} for an overview),
which is based on BRST cohomology 
\cite{BarnichBrandtHenneaux1,BarnichBrandtHenneaux2,BarnichBrandtHenneaux3}, 
has recently yielded important results on 
the classification of allowed first-order deformations 
of the free gauge theory for 
a set of $p$-form fields, $p=1,\ldots,n-1$, in $n\ge 2$ dimensions
\cite{Henneaux2,Henneaux3}.
While this classification analysis is complete for 
massless $p$-form fields with $p\ge 2$
and lists the extended \FT/ and \YM/ types of first-order deformations
for $p\ge 1$,
it did not explicitly treat deformations of massive $p$-form fields
with the mass given by an abelian \CS/ term in the free gauge theory. 
Moreover, integrability conditions 
(\ie/ obstructions to the existence of higher-order deformation terms)
associated with combining the distinct types of allowed 
first-order deformations were not obtained for any $p\ge 1$. 
In the case $p\le 2$, 
these gaps are closed by the deformation results obtained 
in \secref{deformations}. 
In particular, a complete classification of 
first-order geometrical deformations of the free gauge theory 
for a massive/massless set of 1-form and 2-form fields
has been obtained in $n=4$ dimensions, 
including all integrability conditions that arise on such deformations 
(with typical assumptions on the allowed number of derivatives
considered for terms in the gauge symmetries and field equations). 
Also, uniqueness results on deformations to all orders in this setting
have been proved. 
(Interestingly, if the restriction to geometrical deformations is relaxed,
then an additional type of deformation is known to exist 
in the case $p=1$ \cite{Brandt3}.)

There are several directions in which the main results in this paper
could be generalized. 
First, 
an extension of the general massless/massive nonlinear theory 
constructed here for 
\YM/ 1-form gauge fields coupled to \FT/ 2-form gauge fields 
with a \CS/ mass term in four dimensions
is expected to exist in $n$ dimensions,
involving a tower of Lie-algebra valued $p$-form fields $\A{}{(p)}$,
$p=1,\ldots,n-2$, 
with a \YM/ self-coupling on $\A{}{(1)}$, 
a \FT/ self-coupling on $\A{}{(n-2)}$, 
and an extended \FT/ coupling between $\A{}{(1)},\dots,\A{}{(n-2)}$,
in addition to a Higgs coupling of $\A{}{(2)},\dots,\A{}{(n-2)}$
with $\A{}{(1)}$ in the massive case. 

Second, 
it is straightforward to couple such a geometrical nonlinear gauge theory
to gravity. 
In particular, on a spacetime with metric tensor $g$, 
the only structure needed is the Hodge dual operator $*$ determined by $g$,
and the exterior derivative $d$ operator (which is independent of $g$). 
For the case $n=4$ dimensions, 
if the Lagrangian of the nonlinear gauge theory 
given in this paper 
for $\A{}{(1)}$ and $\A{}{(2)}$
is combined with the Einstein gravitational Lagrangian for $g$, 
then this achieves an interesting generalization of 
the Einstein-\YM/ theory
(and there is obvious extension to $n$ dimensions
for $\A{}{(1)},\dots,\A{}{(n-2)}$). 
Of particular interest would be to consider its field theoretic features,
such as black hole solutions, nonabelian monopole solutions, 
and critical behavior in the initial value problem.

\acknowledgements
The author would like to thank the organizers of the
First and Second Workshops on Formal Geometry and Mathematical Physics
where some of the results in this paper were first presented.

\end{document}